\documentclass[10pt,aps,prx,twocolumn,showpacs,notitlepage,floatfix,superscriptaddress,longbibliography]{revtex4-2}
\usepackage{hyperref}
\hypersetup{
colorlinks=false,
colorlinks=true,
citecolor=blue,
linkcolor=blue,
urlcolor=blue}
\usepackage{graphicx}
\usepackage{amsmath}
\usepackage{physics}
\usepackage{amsfonts}
\usepackage{amssymb}
\usepackage{mathtools}
\usepackage{bm}
\usepackage{siunitx}
\usepackage{xfrac}
\usepackage[normalem]{ulem}
\usepackage{orcidlink}  
\usepackage{braket}
\usepackage{cases}

\newcommand{\pcmd}[1]{\textcolor{purple}{#1}}

\DeclareUnicodeCharacter{0229}{\c{e}}

\begin{document}

\title{Improving the efficiency of finite-time memory erasure with potential barrier shaping}
\author{Vipul Rai}
\email{Contact author: d23176@students.iitmandi.ac.in}
\affiliation{School of Physical Sciences, Indian Institute of Technology Mandi, Mandi-175005 (H.P.), India}
\author{Moupriya Das\orcidlink{0000-0003-1851-2162}}
\email{Contact author: moupriya@iitmandi.ac.in}
\affiliation{School of Chemical Sciences, Indian Institute of Technology Mandi, Mandi-175005 (H.P.), India}

\begin{abstract}
Erasure of the binary memory, $0$ or $1$, is an essential step for digital computation as it involves irreversible logic operations. In the classical case, the erasure of a bit of memory is accompanied by the evolution of a minimum amount of heat set by the Landauer bound $k_{B}T \ln2$, which can be achieved in the asymptotic limit. However, the erasure of memory needs to be completed within a finite time for practical and effective computational processes. It is observed that the higher the speed of erasure, the greater the amount of heat released, which leads to unfavorable environmental conditions. Therefore, this is a fundamental challenge to reduce the evolved heat related to finite-time memory erasure. In the present work, we address this crucial aspect in the field of information thermodynamics. We proceed by considering the physical model framework where the two memory states correspond to the two wells of a bistable potential, as in the conventional cases. However, the potential is asymmetric in terms of the width of the two wells. Moreover, the two memory states are separated by a barrier that is asymmetric in structure. This type of asymmetry models the two binary memory states which occupy different phase-space volumes, although they are energetically equivalent, in a general setup. We examine in detail the effect of the degree of asymmetry on the success rate of the erasure process and the work done or heat released associated with it. We find that the asymmetry in the width of the potential wells and the barrier partitioning them, i.e., the two memory states,  plays a very significant role in improving the efficiency of the erasure process, in view of the success rate and the thermodynamic costs. Our thorough simulation study establishes the fact that one can reach below the Landauer bound in an appropriate asymmetric setup. Importantly, it develops a quantitative understanding of the deviation from the Landauer limit as a function of the degree of asymmetry of the potential governing the erasure mechanism. Moreover, through our current work, we identify the effective free energy change for the finite-time bit erasure process as a general lower bound for the work done or evolved heat even when the departure from the Landauer limit is observed. We retrieve the approach towards the Landauer limit in terms of the energetics involved with the erasure mechanism under the symmetric setup.
\end{abstract}

\maketitle

\section{Introduction}

Digital computation has become an inevitable part of our 
everyday life because of its implementation in the 
technological devices that appear to be indispensable to our 
daily needs. These computational processes mainly involve 
logical operations~\cite{Murali_prl_2009,Takeuchi_pre_2018,Shannon_1948} on the binary memories $0$ and $1$ 
to generate outputs. Most of these basic types of logic 
operations, such as OR, AND, XOR, and their negation 
operations, are irreversible in nature~\cite{PhysRevX_Manzano_2024}. This is because 
there are at least two inputs that produce one single output, 
i.e., inputs cannot be determined uniquely 
by observing the output. Information is lost during the 
operational procedure, or it can be interpreted as the 
erasure of memory. Landauer first pointed out that the 
erasure of a classical bit of information is accompanied by an 
unavoidable evolution of heat, which has a lower bound of 
$k_{B}T \ln2$ ~\cite{Landauer_1961,landauer_nature_1998,5391327, bennett1982thermodynamics,Mandal2012,Barbara_pra_2000,chattopadhyay2025landauer,Landi_prl,Chaves_prx_2025,Sander_pre2025}, where $k_{B}$ is the Boltzmann constant 
and $T$ represents the temperature. This change may be 
understood in terms of the decrease of entropy $k_{B}\ln2$ for 
the process that is the result of the conversion of the 
initially available two memory states, to one at the final 
time. As the computational devices form a closed system with 
their surroundings, the overall entropy change remains zero 
and therefore, the reduced entropy for the devices appears 
as a heating effect in the environment, with the value of 
$k_{B}T \ln2$~\cite{Dillenschneider_prl_2009,berut_lutz2012nature,Klaers_prl_2019}. Now, the knowledge of thermodynamics 
suggests that this is the interpretation of the 
amount of the released heat when the process 
is quasi-reversible, i.e., the time for the completion of the 
process is almost infinite. We understand that the 
memory erasure mechanism has to be completed in 
a given amount of time for realistic and effective 
computations. Any finite-time process for 
the erasure of a classical bit of memory would 
result in the evolution of heat, which has a greater value as 
compared to the Landauer bound of $k_{B}T\ln2$~\cite{finite_time_landauer_prl_2022,Proesmans_PhysRevLett_2020,Proesmans_pre,Dago_Ludovic_PhysRevLett_22,Vlatko_RevModPhys.81.1,hong2016experimental,Sagawa_prr2024}. The 
generated heat for the erasure of bits of information 
accumulates depending on the extent of the computational task 
performed on digital devices and released in the environment.  
This becomes an unfavorable situation for the surroundings 
as well as for the devices. Therefore, it is a 
fundamental need to lower the value of the heat released  
for the finite-time memory erasure process and eventually, 
for the practical computations. Consequently, this topic 
emerges as a field of prime significance 
in the domain of information thermodynamics~\cite{Dynamical_Landauer_Principle_prl_2025,thermodynamics_nature_2015,Prx_Esposito_2014,prl_ito_2018,Goold_2016,prx_ciliberto_2017,MARTINI2016108,wolpert2024stochastic}. 

Now, to predict and implement this idea of managing the 
thermodynamic costs associated with the erasure protocols, 
it is required to represent the memory 
by some suitable physical entity and the act of erasing 
has to be mimicked by an apt physical setup. In most of the 
previous studies, the system is considered to correspond 
to a Brownian particle moving in a bistable potential~\cite{berut_lutz2012nature,Dillenschneider_prl_2009,Reguera_pre_2001,Mondal_pre_2010},
which is in a thermal equilibrium with a heat bath. 
The state of the particles in the two separate wells of the 
bistable potential corresponds to two distinct memory states 
$0$ and $1$. To understand the memory erasure in small 
systems, an ensemble of such setups is taken into account.  
Due to the application of a time-linear bias, 
which is considered an erasing force, 
during the time of the protocol, 
one type of memory is erased out of the two. 
The ensemble-averaged value of the work or the heat 
related to this mechanism provides an idea about 
the thermodynamic costs associated with the erasure protocol. 
Some crucial studies have been done to optimize 
the evolved heat associated with the erasure process~\cite{Proesmans_PhysRevLett_2020,Proesmans_pre,Pre_2025_ludovic,Naoto_2018,Mitchison_prl_2020,roldan2014universal_nature,Gopalkrishnan2016,Lee_prl_2022,Saito_prl_2022,Modi_prl_2021,konopik2023fundamental,Sagawa_2025_prr}. 
Different pathways of consideration of the underdamped 
dynamics~\cite{Dago_Ludovic_PhysRevLett_22,Dago_2022,aurell2012refined}, 
time-multiplexed potentials\cite{Talukdar_pre_2017}, final local equilibrium state~\cite{Proesmans_PhysRevLett_2020,Konopik_2020_epl,ciampini2021experimental,Buffoni_pre}, nonequilibrium memory states ~\cite{Konopik_2020_epl,saira_prr_2020,Sagawa_2009,Esposito_2011}, etc. are considered in this regard.   
However, there is still no definite answer to this important 
question of reducing the heat systematically, 
by choosing a particular form of the memory 
or the erasure protocol \cite{Proesmans_PhysRevLett_2020,Proesmans_pre}. In the majority of the 
earlier studies, the dynamics of the Brownian particle whose 
state represents the memory, are governed by 
a symmetric bistable potential, and the two states are 
separated by a symmetric barrier. In the present study, we 
consider the underlying bistable potential to be asymmetric in nature 
with respect to the width of the two potential wells 
and the degree of asymmetry 
can be tuned by an asymmetry parameter~\cite{innerbichler_2020,ciampini2021experimental}. 
The ultimate value of the asymmetry parameter converts 
the potential to a symmetric one. In this framework, 
the memory states are separated by an asymmetric barrier top, 
except for the symmetric form of the potential. 
The chosen asymmetric potential of the 
above-mentioned form can be considered as the general setup 
to model binary memory states when they occupy different 
volumes in phase space; if they are allowed to get thermalized with the heat bath. 
The asymmetry parameter enables tuning of 
the relative phase space volumes of the two memory states. 
However, the two states of the binary memory remain energetically equivalent 
as the energy values of the two minima of the potential, 
representing the two memory states are the same, under the unperturbed condition. 
We perform a detailed study to investigate whether and how 
this applied asymmetry in the potential can improve 
the efficiency of the erasure process, and consequently, 
the performance of the computational tasks. 
The answer is sought in terms of the success rate of the process 
and the thermodynamic costs, i.e., 
the average amount of work-done, evolved heat, and 
free energy related to it. The effect of the general asymmetry 
on the memory erasure procedure has been studied before~\cite{Sagawa_2009,Gavrilov_2016}. 
However, the understanding of the quantitative response 
of the efficacy of the erasure phenomenon 
towards the degree of asymmetry is missing.  
Here, we perform a methodical study 
to interpret how the extent of asymmetry in the potential 
influences the above-mentioned observables 
related to erasure to tune the efficiency of the process. 
The current work indicates a precise path for 
enhancing the effectiveness of the fundamental process 
of memory erasure in a systematic way. 

The paper is organized as follows. In section II, the model 
and dynamics are described. The results have been presented 
and discussed in section III. It is concluded in section IV.
 
\section{Model and dynamics}
We model the erasure process through the dynamics of a Brownian particle
in a bistable potential subject to external forces that induce erasure.
We describe the dynamics using the overdamped Langevin equation:
\begin{equation}
    \gamma \frac{dx}{dt} = -\frac{\partial V(x,t)}{\partial x} + A f(t) + \sqrt{D} \xi(t),
    \label{eq:langevin}
\end{equation}
Here, $x$ represents the position of the Brownian particle. The value of $x$ determines the state of the system. \( \gamma \) is the friction or damping coefficient. 
\( \xi(t) \) denotes Gaussian white noise with statistical properties
\[
\langle \xi(t) \rangle = 0, \quad \langle \xi(t) \xi(t') \rangle = 2 \delta(t - t').
\]
Here, \( D \) is the diffusion coefficient and a measure of the thermal 
fluctuations. It is related to the temperature $T$ and the damping coefficient 
$\gamma$ as \( D = \gamma k_{B}T \), where $k_{B}$ is the Boltzmann constant. 

\begin{figure}[ht]
  \includegraphics[width= 0.95\linewidth,height=4.5cm]{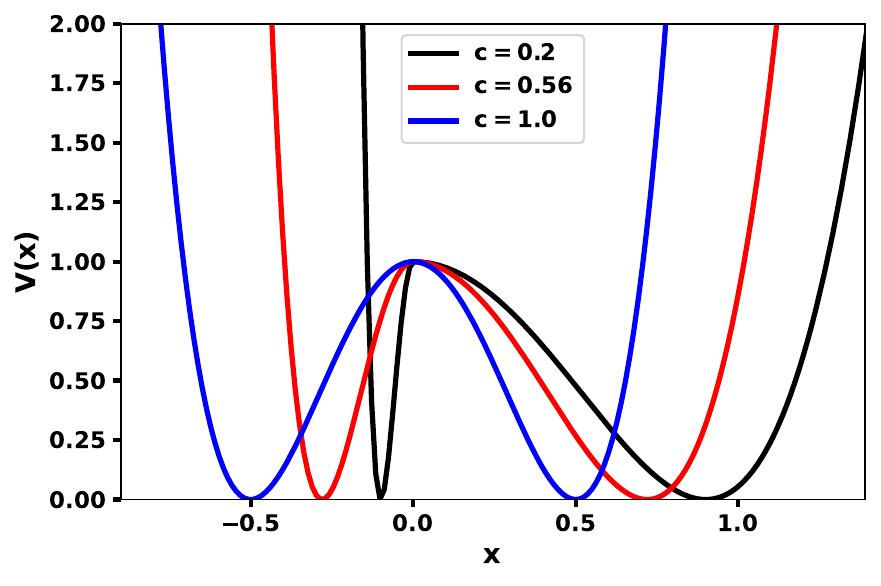}
  \caption{Potential \(V(x)\) with different asymmetry parameter value $c$ for a fixed barrier height of \(h=1\).}
  \label{fig:only_potential}
\end{figure}

\begin{figure*}[ht]

    \includegraphics[width=0.4\linewidth,height=4.5cm]{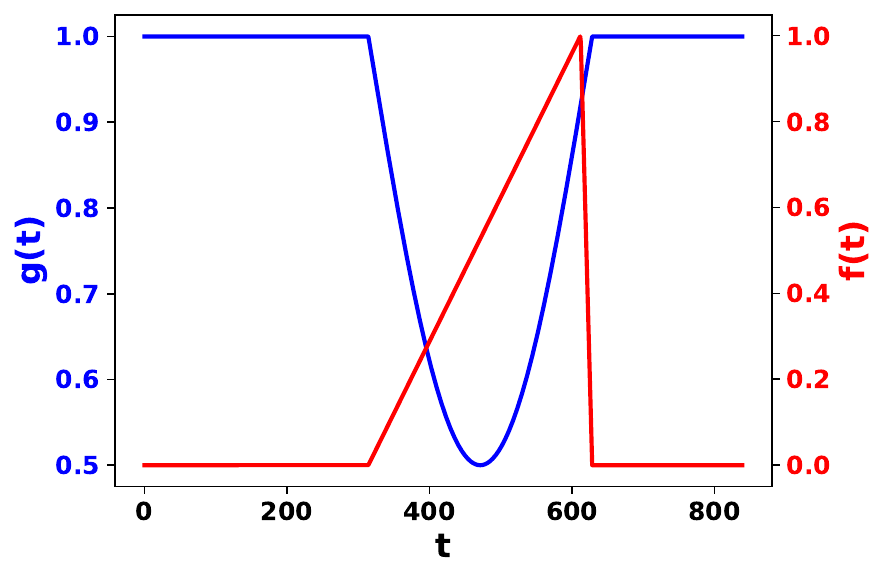}
    \caption[]{The time-series of the tilting force $f(t)$ and the barrier-lowering force $g(t)$.}
  \label{fig:protocol}
\end{figure*}

\begin{figure}[ht]
  \centering
    \centering
    \includegraphics[width=\linewidth, height=4.25cm]{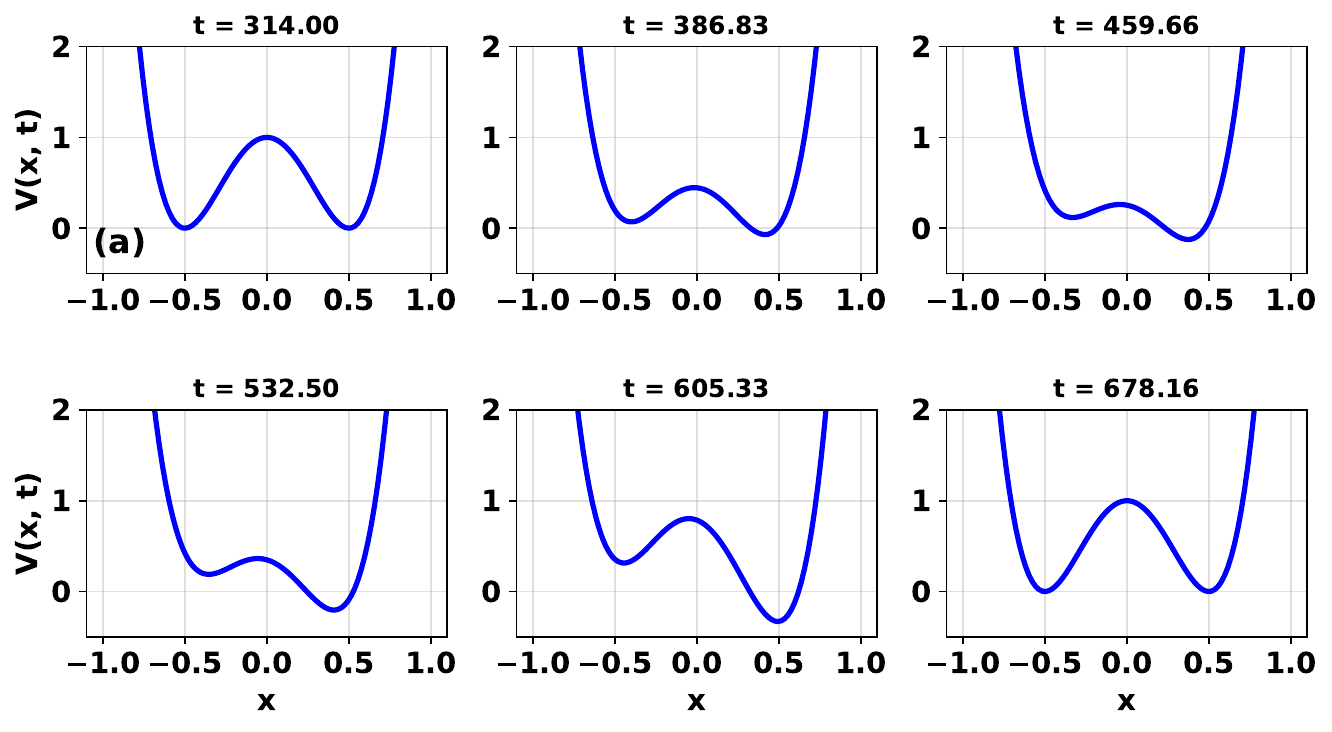}
  \hfill
    \centering
    \includegraphics[width=\linewidth, height=4.25cm]{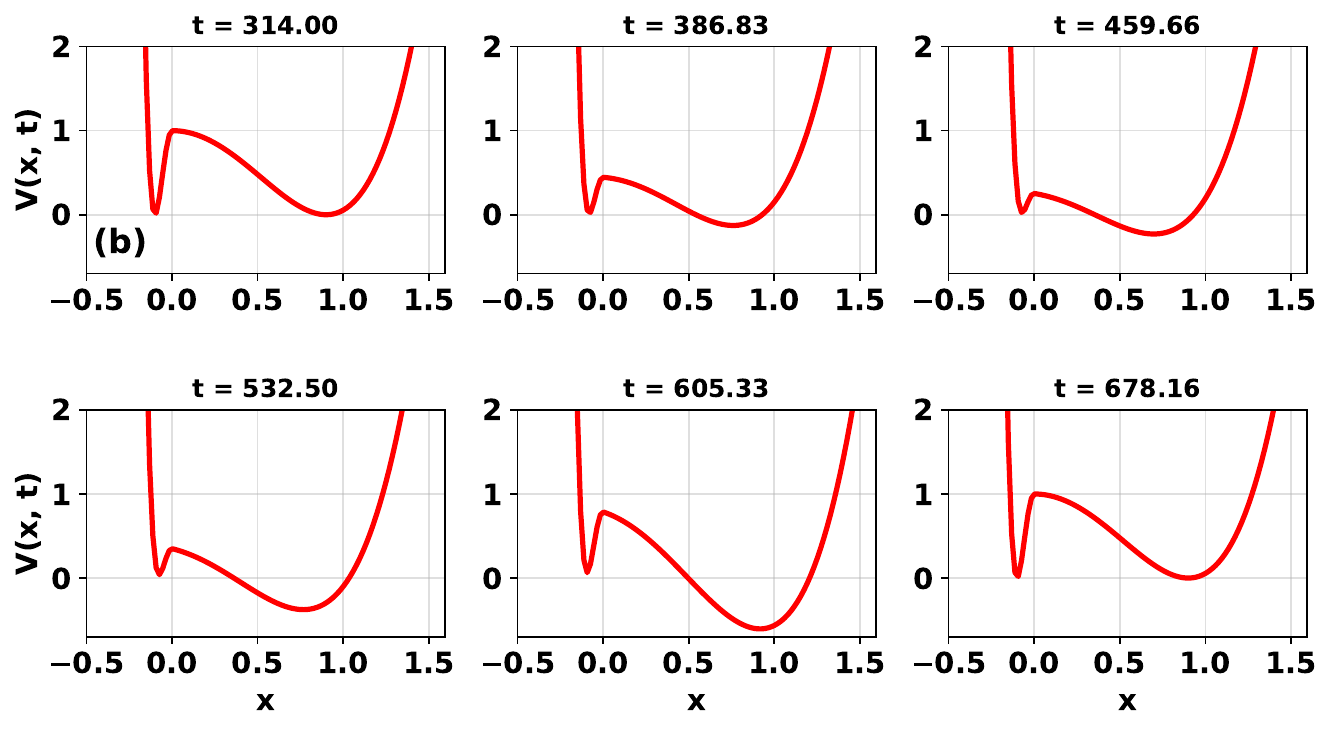}

  \caption{Time evolution of the bistable potential with barrier height $h=1$ during an erasure cycle with $A=0.7$, \(\omega = 0.01\) and \(Z = 0.5\) for the \textbf{(a)} symmetric case \(c = 1\) and the \textbf{(b)} asymmetric case with \(c = 0.2\). Each panel shows the potential landscape at different time points during the erasure protocol.}
  \label{fig:potentials} 
\end{figure}

We employ an asymmetric double-well potential \(V(x,t)\) following the forms used in Refs.~\cite{innerbichler_2020}, where the asymmetry arises from differing well widths. The potential is defined as
\begin{align}\label{2.2}
    V(x,t) &= h \left[ 1 + \left( \frac{2x}{S(x)} \right)^4 - 2g(t) \left( \frac{2x}{S(x)} \right)^2 \right],\\
\textrm{with} \nonumber \\    
    S(x) &= c\, \Theta(-x) + (2 - c)\, \Theta(x),
\end{align}
where \( h \) is the barrier height, \( S(x) \) is a scaling function that regulates the relative widths of the left and right wells, \(\Theta(x)\) denotes the Heaviside function. \(c\) is a positive parameter that introduces asymmetry in the potential by controlling the relative widths of the two wells. Here, \( g(t) \) is a dimensionless, time-dependent function that modulates the barrier height during the erasure protocol. 
The time-independent form of the potential $V(x)$ (corresponding to $g(t)=1$) has been shown in Fig.~\ref{fig:only_potential}. \( A f(t) \) is an external force applied directly to the particle to cause transitions during the process.  
Therefore, the erasure mechanism is characterized by two time-dependent functions: \( g(t) \), the barrier-lowering function, and \( f(t) \), which facilitates the transition of the particle across the barrier, i.e., to the desired memory state, by introducing tilt to the potential. $A$ stands for the amplitude of this tilting force. The mathematical descriptions of these two functions are presented below. The function \(g(t)\), which modulates the height of the potential barrier over time, lowering and raising it within a duration \(t_f = \tau / 2\), can be expressed as, 
\begin{equation}
g(t) = 
\begin{cases}
1 - Z\sin\left( \omega (t - t_0) \right), & \text{if } t_0 \leq t \leq t_0 + t_f, \\
1, & \text{otherwise}.
\end{cases}
\label{eq:barrier_lowering}
\end{equation}
The modulation follows a sinusoidal form with period \(\tau = 2\pi / \omega\) and \(Z\) determines the amplitude of this regulation. The function \(f(t)\) is a sawtooth-shaped tilting force characterized by two time constants, \(\tau_1\) and \(\tau_2\), with the total protocol duration given by \(t_f = \tau_1 + \tau_2\). During the first phase (\(0 < t < \tau_1\)), the external driving force ramps up linearly to a maximum value \(A\), and in the second phase (\(\tau_1 < t < \tau_1 + \tau_2\)), it ramps down linearly to zero. Therefore, $f(t)$ can be represented as follows, 
\begin{equation}
f(t) = 
\begin{cases}
\dfrac{t - t_0}{\tau_1}, & \text{if } t_0 \leq t \leq t_0 + \tau_1, \\
1 - \dfrac{t - t_0 - \tau_1}{\tau_2}, & \text{if } t_0 + \tau_1 \leq t \leq t_0 + \tau_1 + \tau_2, \\
0, & \text{otherwise}.
\end{cases}
\label{eq:tilting}
\end{equation}
At the end of the full erasure cycle, both functions \(g(t)\) and \(f(t)\) return to their initial values (Fig.~\ref{fig:protocol}), and the potential \(V(x)\) is restored to its original form, as illustrated in Fig.~\ref{fig:potentials}. 
The considered form of the barrier-lowering 
and the tilting force satisfies the condition of 
modulation of the barrier height and 
producing the appropriate bias during the erasure mechanism, 
in a simplified setup. Their functional form also fulfils 
the requirement of the return of the final values of the 
forces to their initial values to constitute a cycle. 
Therefore, the above design of the external control 
can be regarded as the elemental drive 
for the erasure process and has been considered 
conventionally, in earlier notable 
studies~\cite{Dillenschneider_prl_2009,berut_lutz2012nature}. 
Next, we proceed to obtain Eq.~\ref{eq:langevin} in dimensionless form to present and deal with a simplified description of the dynamics. The equation has been made dimensionless by scaling the variables and parameters involved in the dynamics with appropriate factors. The dimensionless position variable is expressed as $\tilde{x}=x/l$ where $l$ is the distance between the two minima of the potential and has a numerical value equal to $1$. Time $t$ has been scaled with the factor $t_{l}= \gamma l^{2}/k_{B}T_{R}$ where $T_{R}$ is a reference temperature. The scaling factor $t_{l}$ effectively represents twice the time required for the Brownian particle to travel the distance $l$. The dimensionless time becomes $\tilde{t}=t/ t_{l}$. The amplitude $A$ of the driving force $f(t)$ has been made unitless by dividing by the factor $\gamma l/ \tau$, and the dimensionless quantity has the form of $\tilde{A}=A \tau/ \gamma l$. The frequency $\omega$ appearing in the barrier-lowering term is multiplied by $t_{l}$ to get the scaled frequency as $\tilde{\omega} = \omega t_{l}$.\\
The dimensionless form of the Langevin dynamics can be expressed as,
\begin{equation}
    \frac{d\tilde{x}}{d\tilde{t}} = 16 h \left\{g(\tilde{t}) \frac{\tilde{x}}{S(\tilde{x})^2} -4\frac{\tilde{x}^3}{S(\tilde{x})^4}\right\} + \tilde{A} f(\tilde{t}) + \sqrt{\tilde{D}}\tilde{\xi}(\tilde{t}),
    \label{eq:langevin1}
\end{equation}
where $\tilde{\xi}(\tilde{t})$ is the scaled noise term and $\tilde{D}$ corresponds to the scaled diffusion coefficient which has the form $\tilde{D}=T/T_{R}$. $h$ represents the dimensionless barrier height of the potential.
In the following, we drop the tildes for notational simplicity 
to describe any dimensionless quantities for future reference.

\begin{figure}[ht]
  \centering
  \includegraphics[width= 0.97\linewidth,height=6cm]{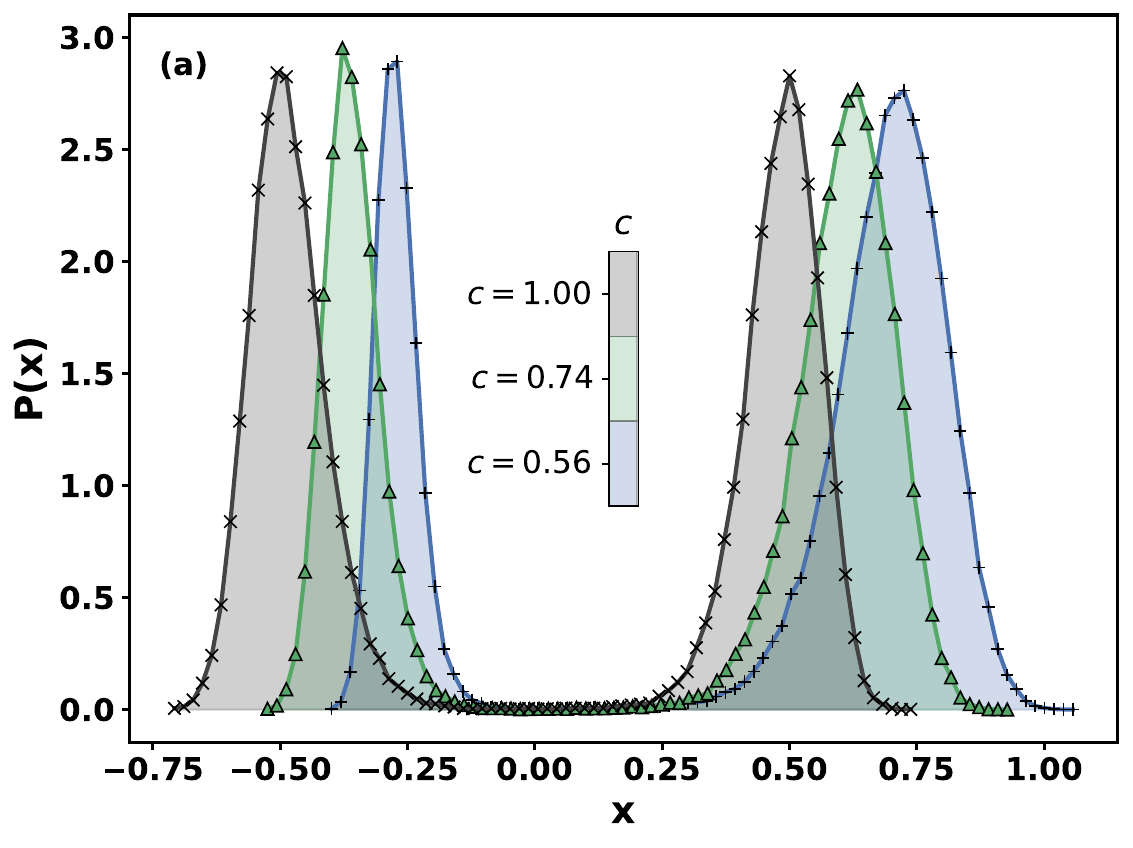}
  \includegraphics[width= 0.97\linewidth,height=6cm]{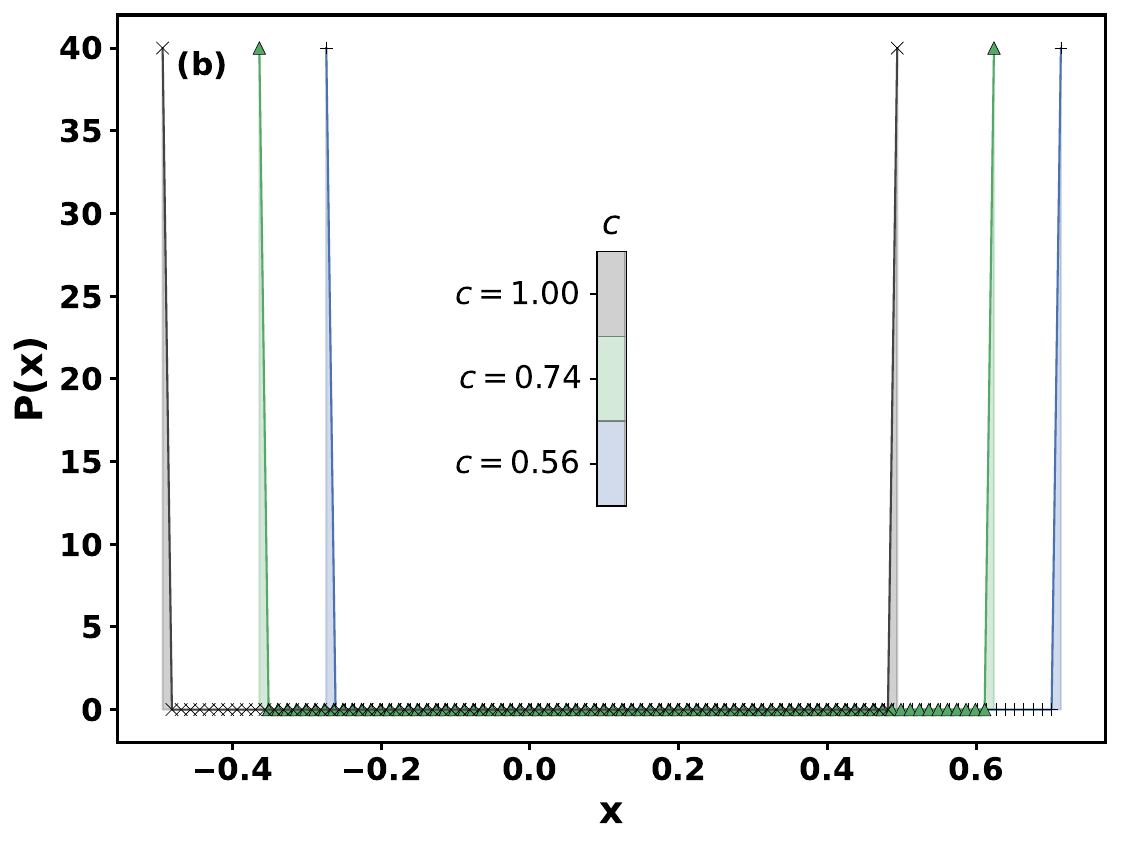}
  \caption{The initial probability distributions of the memory states for three values of $c$; $c=1$, the symmetric case, and $c=0.74$ and $0.56$, two asymmetric cases. (a) Probability distribution of the memory states after thermalization at \(D=0.15\) corresponding to the initial equilibrium condition. The ratios of the memory states, $0$ and $1$, are approximately as follows; $(50:50)$ for $c=1$, $(38:62)$ for $c=0.74$, and $(29:71)$ for $c=0.56$. (b) Initial non-thermalized $(50:50)$ distribution of the memory states for three corresponding values of $c$.}
  \label{fig:position_distribution}
\end{figure}


\begin{figure*}[ht]
  \centering
    \centering
    \includegraphics[width=\linewidth,height=4.6cm]{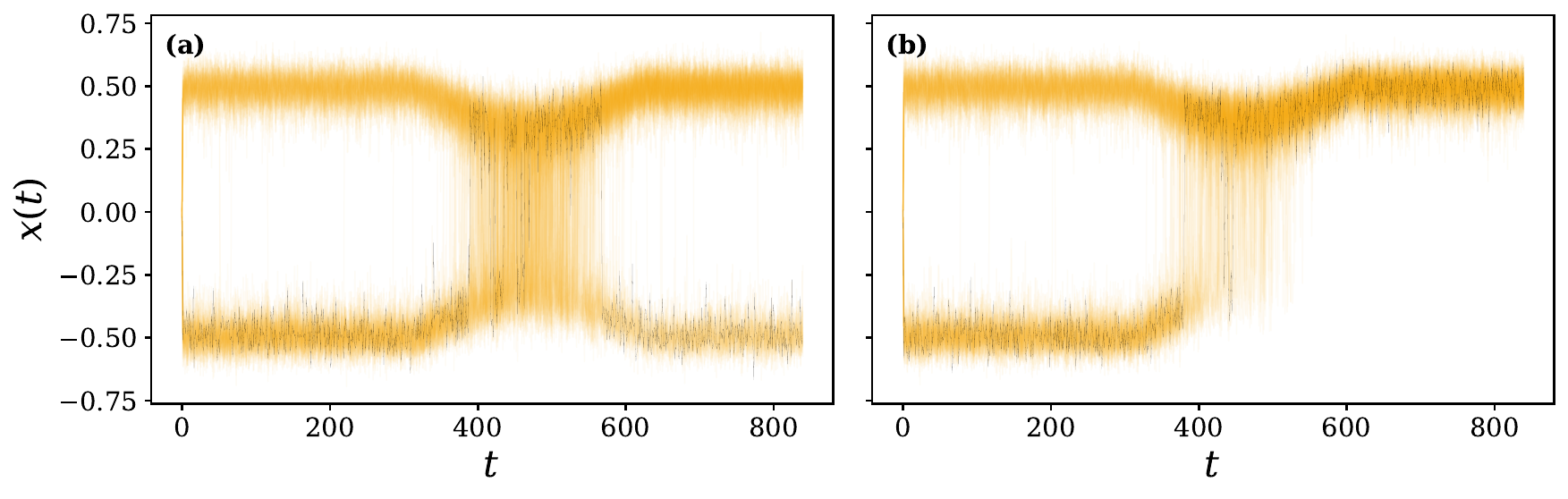} 
    \label{fig:traj_c1}
 
  \vspace{0.35cm}
    \centering
    \includegraphics[width=\linewidth,height=4.6cm]{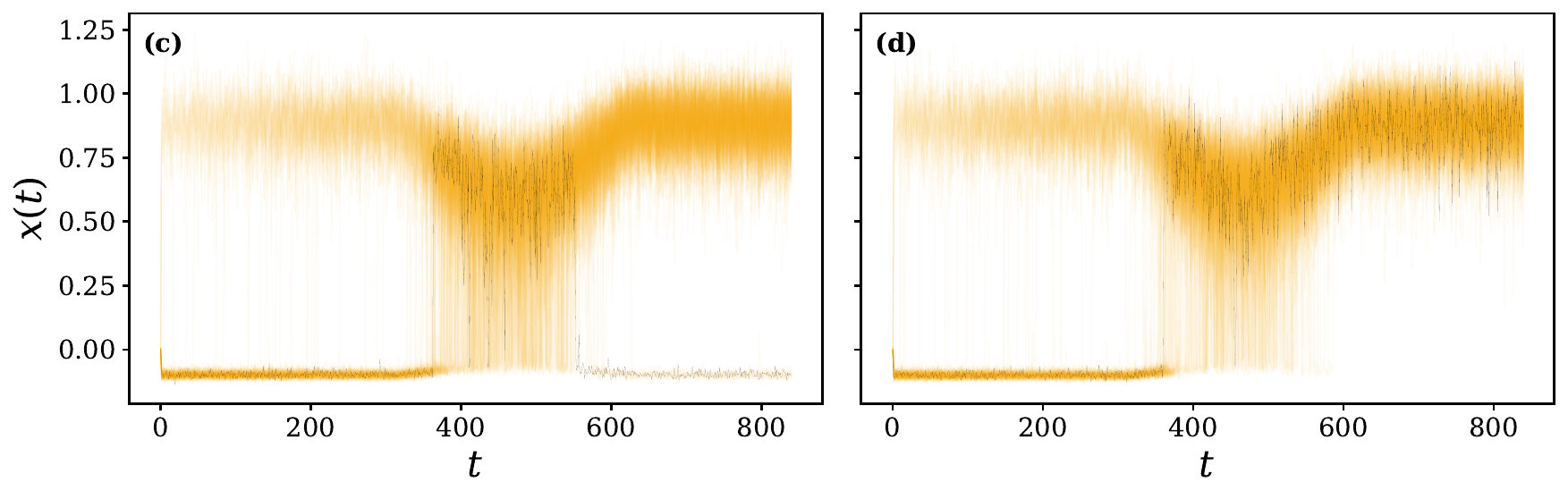}
    \label{fig:traj_c02}
    \vspace{0.35cm}
    \centering
    \includegraphics[width=\linewidth,height=4.6cm]{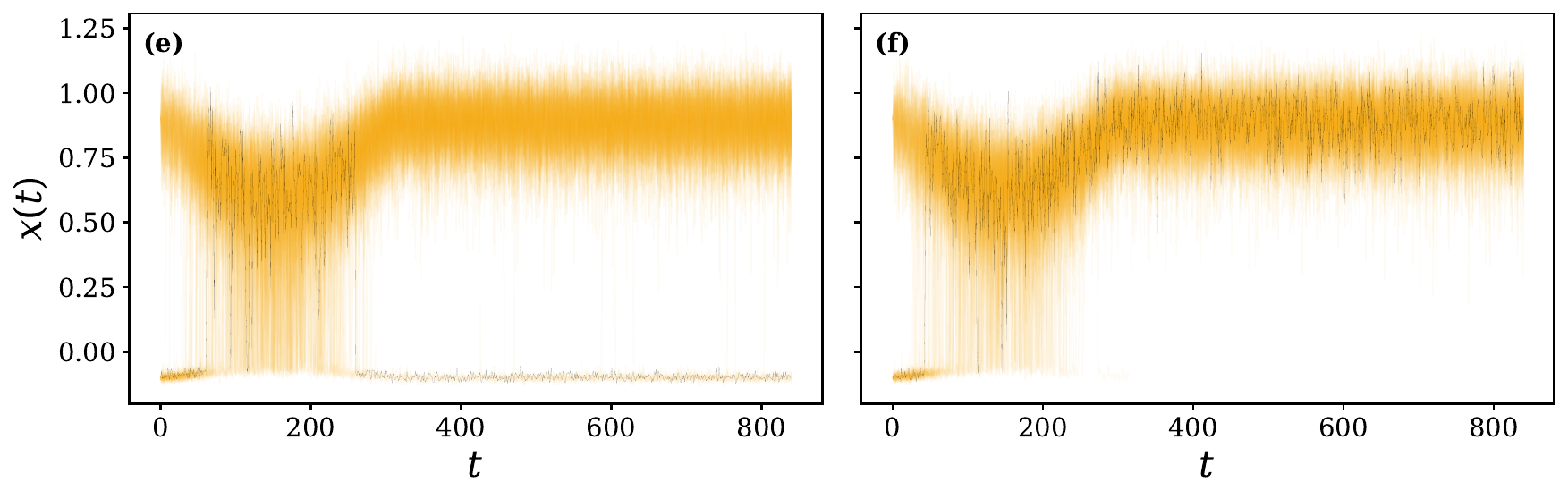}
    \label{fig:trajectories_5050}
    \caption{Stochastic trajectories of 100 particles (shown in orange, with a single trajectory highlighted in black) during the erasure process designed in bistable potentials with barrier height $h = 1$, diffusion constant $D = 0.1$, $\omega = 0.01$, and barrier-lowering depth $Z = 0.5$. 
    \textbf{Top row:} Symmetric potential ($c = 1$), showing \textbf{(a)} incomplete erasure at low tilt $A = 0.2$, and \textbf{(b)} successful erasure at $A = 1$ with equilibrium distribution of initial memory states.  
    \textbf{Middle row:} Asymmetric potential ($c = 0.2$), showing \textbf{(c)} incomplete erasure at $A = 0.1$, and \textbf{(d)} successful erasure at $A = 0.3$ with equilibrium distribution of initial memory states.
    \textbf{Bottom row:} Asymmetric potential ($c = 0.2$), showing \textbf{(e)} incomplete erasure at $A = 0.1$, and \textbf{(f)} successful erasure at $A = 0.3$  with nonequilibrium distribution $(50{:}50)$ of initial memory states. For the cases with initial equilibrium states, the thermalization time is $t_{0}=314$. The ramp-up and ramp-down times for the tilting force are $\tau_{1}=298$ and $\tau_{2}=16$, respectively, for all cases.}
  \label{fig:trajectories}
\end{figure*}

\section{Results and discussions}

In the classical scenario, the Landauer bound is understood starting from an 
initial equal distribution to the memory states, $0$ and $1$, and then ultimate 
conversion to one unique memory state. In terms of the physical system, it implies 
initiating the process with equal occupancy of the Brownian particles 
in both wells, and then transferring all the particles through the erasure mechanism 
to a given well, which represents the final target memory state. 
Now, in the case of the symmetric bistable potential, 
this initial state is also the equilibrium state 
of the system before the start of the erasure protocol. 
Therefore, we can say that the Landauer limit is realized starting with the 
equilibrium memory. This initial equilibrium distribution can be achieved 
if we allow all Brownian trajectories to get adjusted with the thermal bath 
for a sufficiently long time~\cite{Aurell_2011_PhysRevLett,Dillenschneider_prl_2009,Esposito_2011,Patrick_pre}. During this time, an individual Brownian particle gets settled in either of the two wells in the absence of any external force. When we consider an ensemble of such particles, the occupancy in the two wells will follow an equilibrium, more precisely, a steady-state distribution, in the long-time limit. Now, there is one important point to note. Although after this thermalization period, an equal equilibrium distribution in both the memory states is attained in the case of the symmetric potential, the thermally equilibrated distribution remains different for the asymmetric potentials. It also varies depending on the degree of asymmetry in the form of the potential. This is because after this equilibration time, the distribution follows from the steady state probability of occupancy of the two wells, which is now unequal for different wells due to their disparate widths in the asymmetric cases. 
Also, the thermalized distribution depends on the extent of asymmetry in the potential as the asymmetry parameter $c$ governs the relative widths of the two wells, thereby influencing the steady-state probability of their occupancy. Therefore, we can say that our choice of the potential of the asymmetric forms leads to the realization of the erasing process, starting from a condition when the two memory states occupy different phase space volumes, if we allow thermalization. However, the two memory states possess similar energy, initially, as the two minima of the potential correspond to the same energy value before the start of the erasure process. Consequently, we indicate that we start the erasing protocol with distributions of the memory in both the states, $0$ and $1$, after the thermalization period. However, the corresponding occupancies of these two states are characteristics of the underlying potential. Therefore, one of the distinct features of our present study is that when we carry out the erasure process with a starting condition of equilibrium memory, the initial distribution of the memory states is unique based on the extent of asymmetry of the governing potential. We have presented representative initial equilibrium memory distributions for three conditions, one for the symmetric case $c=1$ and two asymmetric forms of the potential with $c=0.74$ and $0.56$ in Fig.~\ref{fig:position_distribution}(a). We perform our investigations on the erasure phenomenon with these initial conditions, for a range of asymmetric potential forms, to understand the effect of the asymmetry on it when we start with the equilibrium memory states. 

We infer that it is also important to analyze the effect of the asymmetry in results for the erasure mechanism when we start from the equally distributed initial memory states. This is because consideration of that initial state sets the limit of $k_{B}T \ln2$ for the erasure of a classical bit of memory and interprets the Landauer bound. Therefore, we also carry out a separate set of investigations starting with the same condition of the equal initial distribution to the two memory states, irrespective of the structure of the potential. For the present case, we do not allow any thermalization time before the start of the protocol. We realize that this distribution will not conform to the thermalized distribution of the systems without any bias for the asymmetric cases, and can be regarded as the nonequilibrium initial condition of the memory states for those systems. Some representative probability distributions for the initial conditions with nonequilibrium memories have been illustrated in Fig.\ref{fig:position_distribution}(b). In this plot, the $c$ values for the potential have been chosen as the same as those considered in Fig.\ref{fig:position_distribution}(a) for the initial equilibrium condition scenarios. Performing the studies with these initial nonequilibrium distributions of memory states also provides us with the scope of comparisons of the findings between the cases of initial equilibrium and nonequilibrium memory conditions. 


\subsection{Success rate}

First, we analyze the efficiency of the erasure mechanism concerning the success rate of the process, which is the ratio of the successful erasure cycles and the total number of cycles for the ensemble. To proceed, we have simulated the overdamped Langevin equation~\eqref{eq:langevin1} using an improved Euler algorithm~\cite{press1992numerical}. Gaussian white noise \( \xi(t) \) is generated using the Box–Muller algorithm. First, we consider the case of starting the erasure protocol with equilibrium memory. In these simulations, at the start of the time, we place all the particles of the ensemble at position \( x = 0 \), i.e., at the barrier top, and let the system get thermalized with the heat bath at temperature \(T\) and arrive at their steady-state distribution according to the structure of the underlying potential. As we consider a range of asymmetric forms of the potential, we get unique steady-state distributions in two wells after the initial thermalization period, as discussed above. We have examined the attainment of this steady-state distribution for all cases studied.\pcmd{To express the scenario in terms of the representation of the memory states, we can say that both the memory states get occupied after the thermalization period, following a steady-state distribution specific to the characteristic asymmetry of the governing potential. For a symmetric potential, this leads to an equal population of both memory states, ``0'' and ``1'' just before the start of the erasure protocol.}

In the case of the erasure of binary memories, one of the memory states vanishes at the end of the process, while initially, both memory states are present. This is an essential step of the irreversible logic operations, in which many (at least two) input memory states are mapped onto one output memory state, thereby erasing information of the starting condition. In a physical setup, this mechanism is modeled in a bistable potential where the states of an ensemble of Brownian particles represent the binary memory $0$ and $1$. The erasure force is applied to the particles, which are initially distributed in both wells, so as to drive all the particles in the target well or to the desired memory state at the end of the protocol. As the particles in the ensemble are indistinguishable, the information about the starting position for each particle is lost during the erasure mechanism. 
When the barrier height of the potential is significantly greater than the thermal energy \( k_{B}T \), the particle remains confined in one of the potential wells, i.e., at one of the memory states, and is unable to transition between them. To make the passage happen to the target well, if the particle is initially at the other well, we use a process where we lower the barrier height and apply a tilting force at the same time after the particle has had time to relax.~\cite{Dillenschneider_prl_2009,Jun_prl_2014,PhysRevResearch_Jun}. In our applied protocol, we aim for the particles to reside in the right well at the end of the process. The tilting force is applied accordingly in the appropriate direction. Therefore, following this mechanism, the goal is to erase the memory state $0$, so that ultimately, the memory state $1$ is occupied, starting from an initial steady-state distribution where both the memory states exist. 

To examine the behavior of the system under the applied protocol, we consider an ensemble of 100 particles and let the system evolve according to the mechanism described above using a time step of \( \Delta t = 10^{-4} \). We carry out the simulations with a very small time-step value, as the barriers are very steep from the narrower side of the asymmetric potentials. This has been done to ensure the accuracy of the solutions of the dynamics and calculations of the concerned quantities following it. We notice that after the thermalization period of duration \( t_0 \), the particles in the ensemble typically settle near either the left or right well minimum, i.e., around $-c/2$ or $(2-c)/2$. As mentioned above, for a symmetric potential, the probability of finding a particle in either well is equal (i.e., one-half) at this point in time. However, in the asymmetric case, this steady-state probability is biased and depends on the relative widths of the two wells. Then, we examine the behavior of the trajectories after the erasure forces are applied to the system. The completeness of the erasure process is understood as follows. If the $x$ values for the ensemble centres around the right well minimum, i.e., near $(2-c)/2$ at the end of the erasure protocol, the process is regarded as complete. This is because we consider the right well or the memory state $1$ to be our target state. The process remains incomplete if we have a distribution of trajectories in both the wells or the memory states at the end of the erasure cycle. We observe that the erasure process is incomplete if the amplitude \((A)\) of the tilting force is below a certain threshold. In contrast, when \( A \) exceeds or equals this threshold value, complete erasure is achieved in both the symmetric and asymmetric cases. The results have been shown in Figs.~\ref{fig:trajectories}{\textcolor{blue}{(a)-(d)}}. 

Next, we analyze the case where particles are initially distributed equally in both wells or memory states and do not allow thermalization. This condition corresponds to a nonequilibrium situation for the initial memory states in the asymmetric potentials. The cases of incomplete and complete erasure for this particular initial condition are shown in Figs.~\ref{fig:trajectories}{\textcolor{blue}{(e)-(f)}}. We infer from the above observations that complete erasure can be achieved both for symmetric and asymmetric cases, and for different initial situations with equilibrium and nonequilibrium memory. However, the constraint in terms of the amplitude of the tilting force to achieve full erasure varies in different cases. 

\begin{figure*}[ht]
    \centering
    \includegraphics[width=0.9\linewidth,height=6cm]{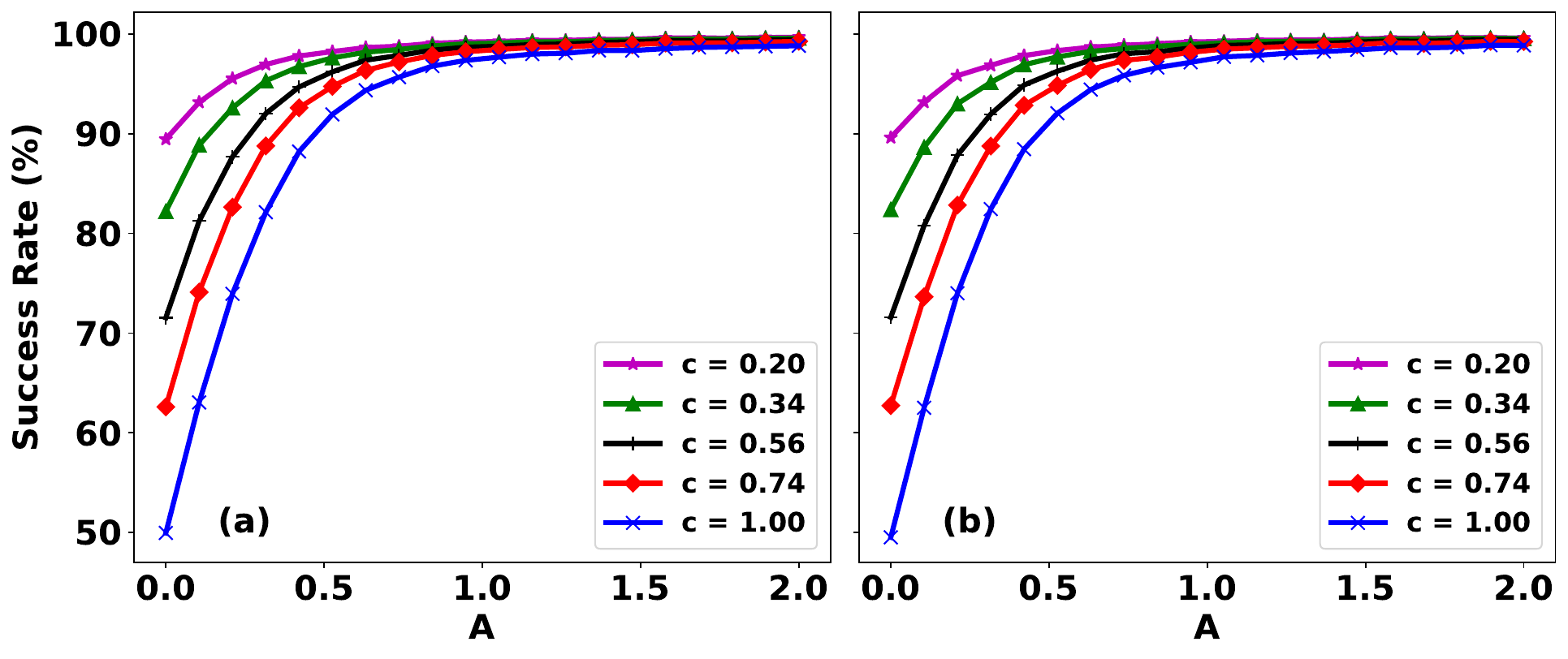}
    \caption{The erasure performance as a function of the driving amplitude of the tilting force $A$ is illustrated. Panels (a) and (b) show how the success rate changes with \( A \) for different asymmetry parameters \( c \). In both panels, barrier height \( h = 1 \), noise strength \( D = 0.15 \), the barrier-lowering depth is \( Z = 0.5 \), \( \omega = 0.01 \), \( \tau_1 = 298 \), and \( \tau_2 = 16 \). In panel (a), the results are for the erasure mechanisms when the particles are initially positioned at $x=0$ at the start of the simulations. Then they are allowed to get thermalized with the heat bath for the duration $t_{0}=314$, whereas for the data in panel (b), the particles have an initial 50-50 distribution without thermalization. For both cases, the ramp-up and ramp-down times for the tilting force are $\tau_{1}=298$ and $\tau_{2}=16$, respectively.}
    \label{fig:erasure_0_position_D0.15}
\end{figure*}

To quantify the above understanding, we plot the success rate of the erasure process, defined as the fraction of particles that occupy the target state, i.e., the memory state 1, represented by the right well of the potential, by the end of the protocol, as a function of the driving amplitude \( ( A )\). The results are obtained from an ensemble of \( 10^5 \) particles. We observe that in the presence of greater asymmetry, a significantly smaller external driving force is sufficient to achieve a 100\% success rate, in contrast to the symmetric case, as shown in Fig.~\ref{fig:erasure_0_position_D0.15}{\textcolor{blue}{(a)-(b)}} for two different initial conditions starting with the equilibrium and nonequilibrium memory states, respectively. In both of the situations, we get similar observations.

\begin{figure*}[ht]
  \centering
  \includegraphics[width= 0.9\linewidth,height=6cm]{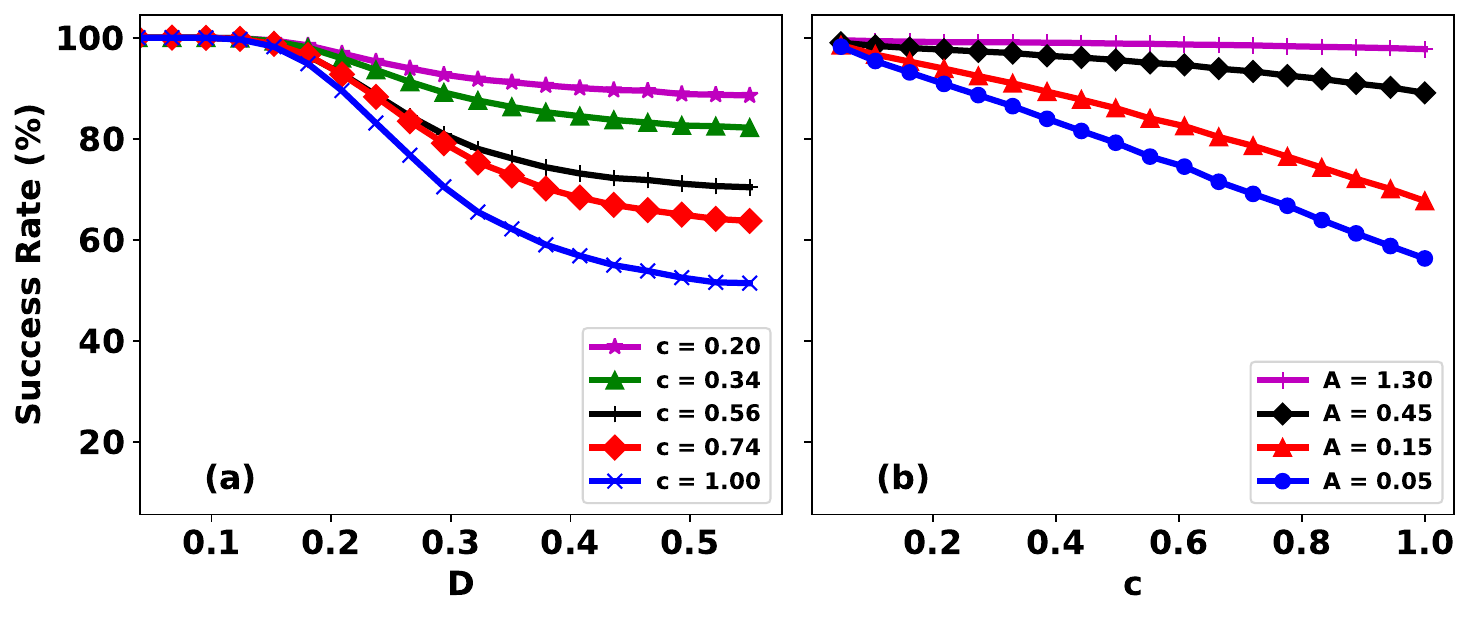}
 \caption{
Panel (a) shows the success rate as a function of the noise strength \( D \), for different asymmetry parameter values with fixed barrier height \( h = 1 \), tilting force amplitude \( A = 1.50 \), protocol shape parameter \( Z = 0.5 \) and $\omega = 0.01$. Panel (b) presents the success rate as a function of the asymmetry parameter \( c \), for fixed values of the barrier height \( h = 1 \), noise strength \( D = 0.15 \), protocol shape parameter \( Z = 0.5 \), and $\omega = 0.01$. For the results in both panels, the particles are placed at $x=0$ at the start of the simulations and are allowed to get thermalized with the heat bath. The thermalization time is \( t_0 = 314 \), and the ramp-up and ramp-down durations of the tilting force are \( \tau_1 = 298 \) and \( \tau_2 = 16 \), respectively.}  
\label{fig:e_d_c_0_position}
\end{figure*}

Next, we have shown how the success rate changes with another important quantity influencing the dynamics, the noise strength \( D \) in Fig.~\ref{fig:e_d_c_0_position}{\textcolor{blue}{(a)}}, while keeping the external force constant at \( A = 1.50\), for different extent of asymmetries in the intrinsic potential. We detect that as the noise strength increases, the success rate decreases, reaching a saturation point in the higher range of $D$. This behavior at higher noise strength can be attributed to spontaneous thermal fluctuations that increasingly drive the particles to the wrong well, or the memory state, thereby reducing the effectiveness of the erasure. 

We understand from the previous results of the variation of the success rate or the success rate as a function of the tilting force amplitude $A$ and the noise strength $D$, for different values of the asymmetry parameter $c$, that the success rate is highly sensitive to the degree of asymmetry present in the potential. Therefore, we perform a more direct study to observe the functional dependence of the success rate on the asymmetry parameter $c$. This explicit variation of the success rate against $c$ is plotted in Fig.~\ref{fig:e_d_c_0_position}{\textcolor{blue}{(b)}}, where we show how the success rate changes with the asymmetry parameter \( c \) for different levels of the external driving force amplitude \( A \) for a fixed value of noise strength $D$. For lower values of \( A \), we observe a higher success rate in the strongly asymmetric regime, but as the system approaches symmetry (\( c \to 1 \)), the success rate drops significantly. However, for sufficiently high external forces (e.g., $A = 1.30$), the success rate remains high (around 95\%) even in the symmetric case ($c = 1$) and does not vary substantially as the asymmetry is reduced. This observation suggests that the effect of asymmetry becomes more prominent when the amplitude of the tilting force is less. This further justifies the fact that the asymmetry factor in the potential makes the transition favorable, which was less probable due to the impact of the erasure protocol alone.

\begin{figure}[ht]
  \centering
  \includegraphics[width= 0.94\linewidth,height=5.5cm]{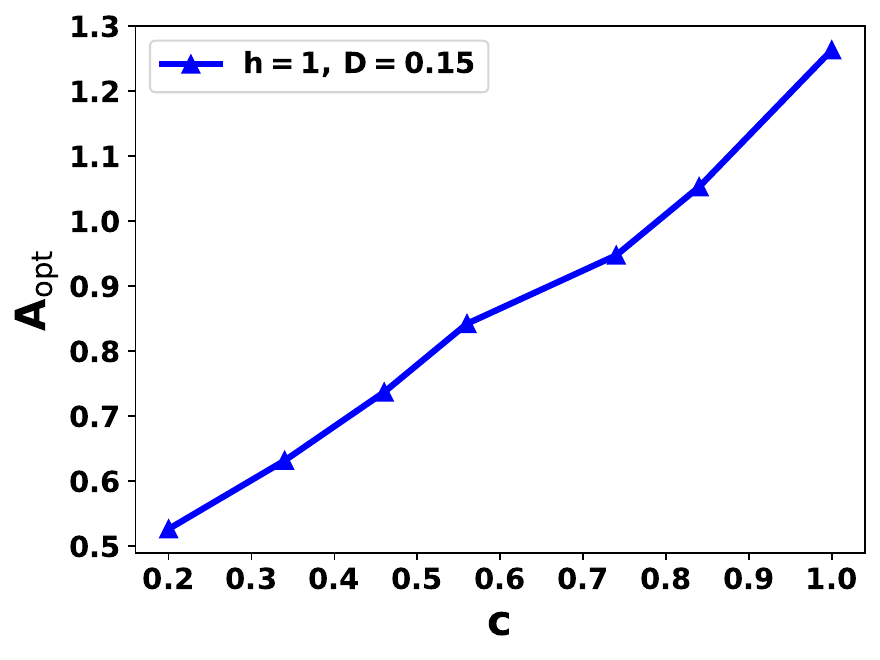}
 \caption{
Optimal driving amplitude \(A_{\text{opt}}\) as a function of the asymmetry parameter \(c\), shown for a fixed barrier height of $h=1$ and noise strength \(D = 0.15\). 
}

  \label{fig:table_plot_combined_o_position_D_0.15}
\end{figure}

Further, we determine an optimal driving amplitude $A_{\textrm{opt}}$ where over $98\%$ erasure occurs for different asymmetries. The data for a fixed barrier height $h=1$ and noise strength $D=0.15$ have been recorded in TABLE~\ref{tab:opt_amp_vs_c}. The results have been presented graphically in Fig.~\ref{fig:table_plot_combined_o_position_D_0.15}. We detect that the value of $A_{\textrm{opt}}$ is much smaller for the asymmetric cases than for the symmetric one. These data provide a systematic representation of our previous understanding that less tilting force is required to make a successful erasure when the intrinsic potential bears asymmetry.

\begin{table}[htb]
\centering
\begingroup
\normalsize  
\setlength{\tabcolsep}{6pt} 
\renewcommand{\arraystretch}{1.4} 
\begin{tabular}{c|ccccccc}
$c$ & 0.20 & 0.34 & 0.46  & 0.56  & 0.74 & 0.84 & 1.00\\
\hline
$A_{\text{opt}}$ & 0.53 & 0.63 & 0.74 & 0.84 & 0.95 & 1.05  & 1.26 \\
\end{tabular}
\caption{Optimal amplitude $A_{\text{opt}}$ as a function of the asymmetry parameter $c$. For \(h=1\) and \(D=0.15\)}
\label{tab:opt_amp_vs_c}
\endgroup
\end{table}

Collectively, these results suggest that the Brownian particle prefers the higher entropy state. In the asymmetric setup, the right well corresponds to a state with higher entropy. As according to the erasure mechanism considered here, the final occupancy of the right well, i.e., the memory state $1$, is regarded as the successful process; the designed asymmetric setup facilitates the process of erasure.

\begin{figure}[ht]
  \centering
  \includegraphics[width= 0.95\linewidth,height=10cm]{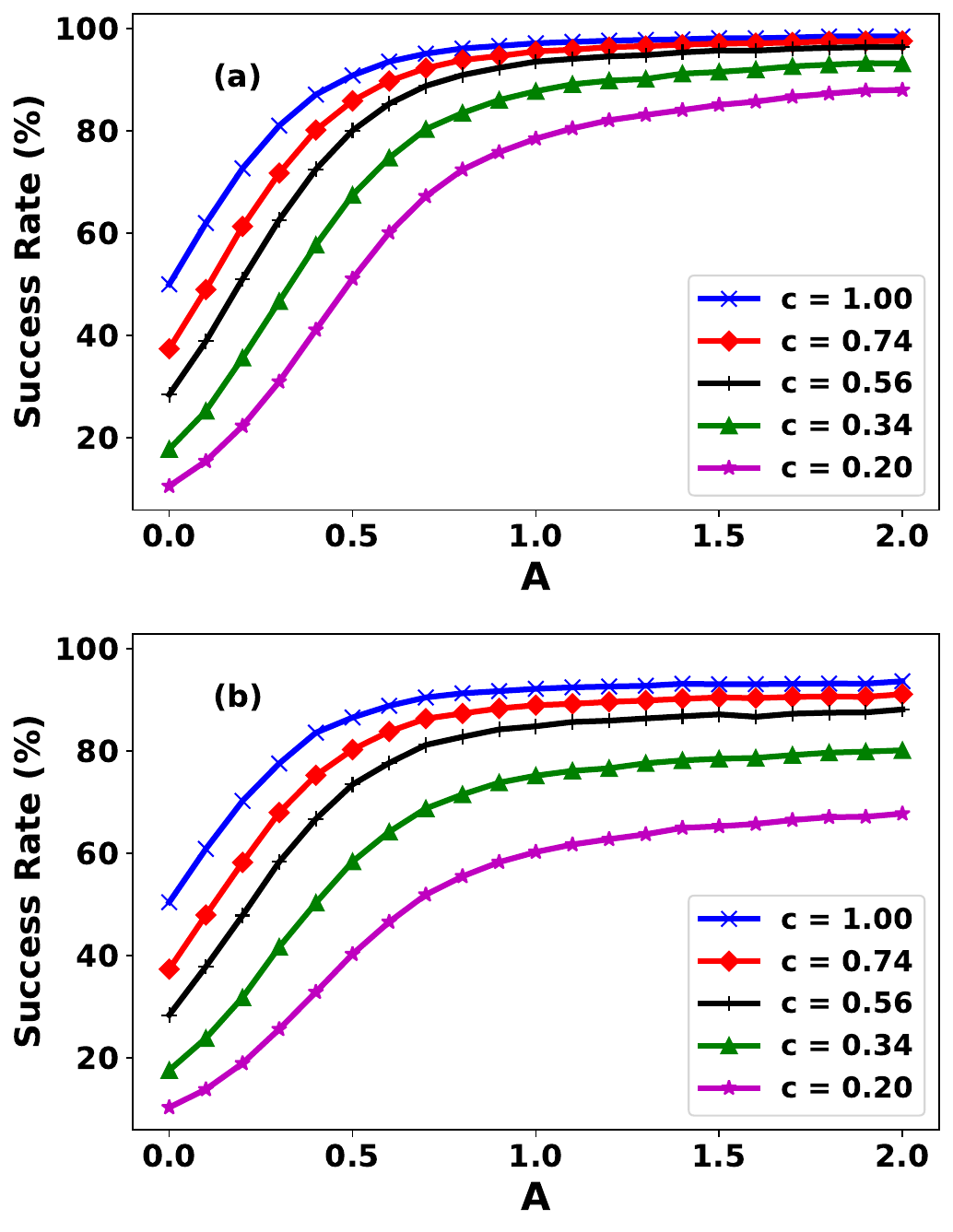}
 \caption{
The success rate as a function of the tilting force amplitude $A$ for the reverse mechanism for different asymmetry parameters $c$. The reverse protocol refers to the scenario where, after the successful erasure, the particle arrives at the left well, i.e., at the memory state $0$. The parameters are fixed at \( h = 1 \), \( D = 0.15 \), $z=0.5$, and $\omega = 0.01$. The results have been presented for the protocols with initial (a) equilibrium and (b) 50:50 distribution of memory states. For case (a), the system undergoes thermalization for a duration of \( t_0 = 314 \). No thermalization period is allowed for case (b). In both situations, the protocol phase consists of the ramp-up and ramp-down durations of \( \tau_1 = 298 \) and \( \tau_2 = 16 \), respectively, for the tilting force.
}

  \label{fig:reverse_0D_0.15_1}
\end{figure}

\textit{Erasing from memory-state ``1'' to memory-state ``0''}: To further confirm this explanation, we also performed the reverse erasing mechanism, where finally the left well or the memory state $0$ is expected to be occupied. The variation of the success rate or the success rate as a function of the tilting amplitude $A$ is plotted in Figs.~\ref{fig:reverse_0D_0.15_1}\textcolor{blue!}{(a)-(b)}, for the systems with initial equilibrium and 50:50 distributions of memory states, respectively. In this case of the reverse protocol, it is observed that a larger tilting force is required to perform the erasure when the system has asymmetry. This again supports our statement concerning the greater preference for the occupancy of the higher entropy state.

In all of the above-mentioned investigations, we verify the findings based on the initial equilibrium and nonequilibrium conditions. Our studies reveal that the observations are similar for both cases. These facts point towards the generality of the effect of the asymmetry on the success rate, irrespective of starting the protocol with equilibrium or nonequilibrium memory states.



\subsection{Heat and work}
Next, we focus on the essential thermodynamic quantities required to describe and interpret the efficiency of the erasure process from the perspective of energetics. These quantities are mainly the work done $W$ and the evolved heat $Q$ associated with the erasure mechanism. We proceed to estimate these observables by employing the fundamental relations of thermodynamics. We implement the specific characteristics of the system and dynamics for our present study in these relations while doing the measurements. The change in the internal energy or the total potential energy is another fundamental energy function that contributes to the development of a detailed description of the energetic aspects of the erasure process. It is also necessary to depict the connection between the work and heat involvement for the protocol. The change in the total potential energy of the Brownian particle during the erasure process is defined as~\cite{Sekimoto_1998,udo_2012}, 
\begin{equation}\label{3.1}
\Delta U = U(x(t_0 + t_f), t_0 + t_f) - U(x(t_0), t_0),  
\end{equation}
where \(U(x, t) = V(x, t) - xA f(t)\) represents the intrinsic potential combined with the potential appearing due to the external driving force. $V(x,t)$ can have an asymmetric or a symmetric form depending on the value of the parameter $c$ present in its expression, as discussed before. The total work $W$ performed on the system during the erasure process can be calculated considering the change in the internal energy as~\cite{Moupriya_PhysRevResearch,udo_2008},
\begin{equation}\label{3.3}
    W\,  = \int_{t_0}^{t_0 + \tau_f} \, \, \frac{\partial U(x,t)}{\partial t} \,  dt.
\end{equation}
For the specific protocol considered in our present study, $W$ can be written in a more explicit form as, 
\begin{equation}\label{3.3b}
 W =\int_{t_0}^{t_0 + \tau_f} \, \, \frac{\partial V(x,t)}{\partial t} \,  dt \,  - \int_{t_0}^{t_0 + \tau_f} \, \, x\, A \dot{f} \,  dt
\end{equation}
We estimate $\Delta U$ and $W$ for each trajectory using the numerical solution of $x$ according to the Langevin dynamics (Eq.~(\ref{eq:langevin})). These two thermodynamic functions are calculated by employing Eq.~(\ref{3.1}) and Eq.~(\ref{3.3b}), respectively. 
According to the first law of thermodynamics, the heat dissipated into the thermal bath \(Q = \Delta U - W \). Therefore, $Q$ can be measured from the solutions of Eq.~(\ref{3.1}) and Eq.~(\ref{3.3b}). As we are dealing with small systems subject to thermal fluctuations, all these quantities, $\Delta U$, $W$, and $Q$, are stochastic variables for a single trajectory. Their ensemble-averaged values, $\langle \Delta U \rangle$, $\langle W \rangle$, and $\langle Q \rangle$ are considered the reliable statistical measures of the quantities and are referred to for further discussions. Also, the first law of thermodynamics takes the form as $\langle \Delta U \rangle = \langle Q \rangle + \langle W \rangle $ for small systems \cite{peliti2021stochastic,H_Kantz,toyabe2010experimental,hasegawa2010generalization,Ueda_2014_pre,Supriya_prr_2024,RevModPhys.97.015002}.    
Since we have a periodic driving force that satisfies \(f(t_0 + t_f) = f(t_0) = 0\), the net contribution of the external drive vanishes over the duration of a period. During a complete erasure cycle, the average change in potential energy remains unaltered, that is, \(\langle \Delta U \rangle = 0\). Therefore, the mean work performed on the system equals the heat dissipated into the bath: \( \vert \langle W \rangle \vert = \vert \langle Q \rangle \vert \).

\begin{figure*}[ht]
  \centering

    \centering
    \includegraphics[width=\linewidth, height=5cm]{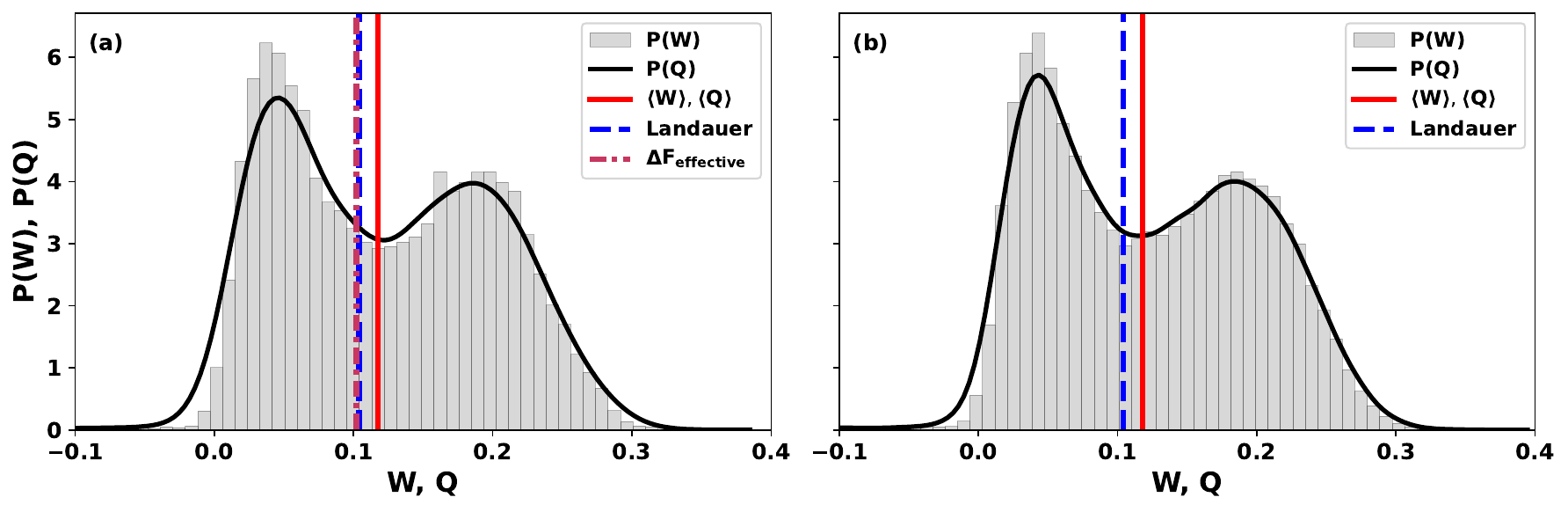}

    \centering
    \includegraphics[width=\linewidth, height=5cm]{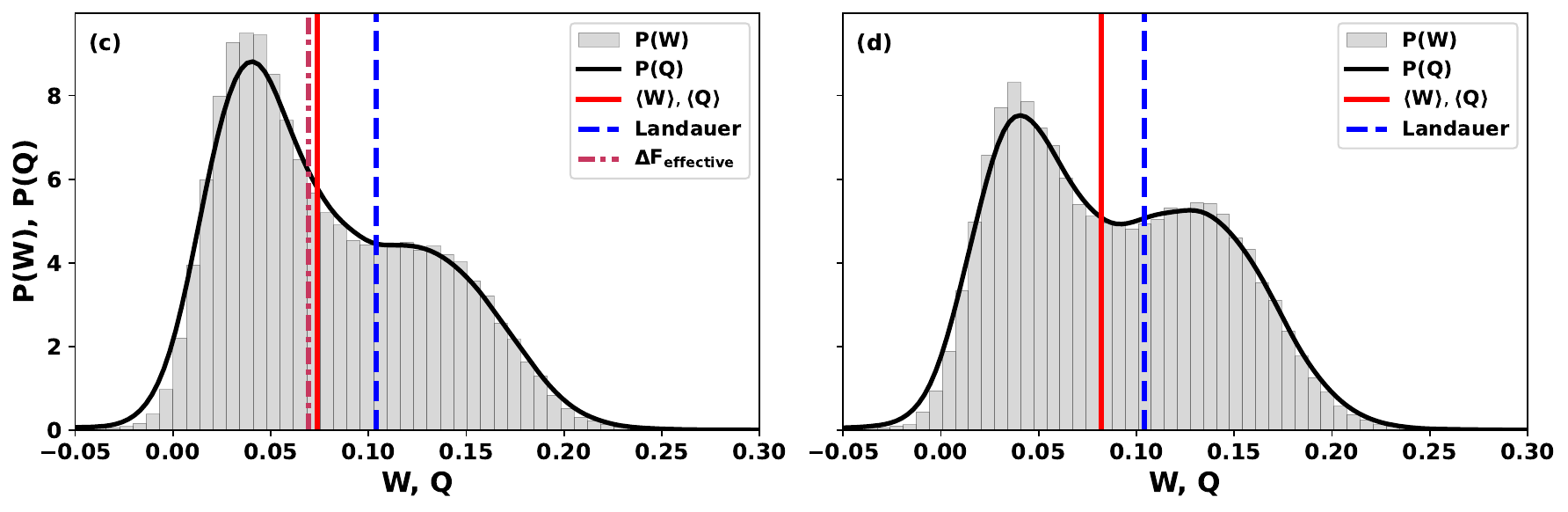}


    \centering
    \includegraphics[width=\linewidth, height=5cm]{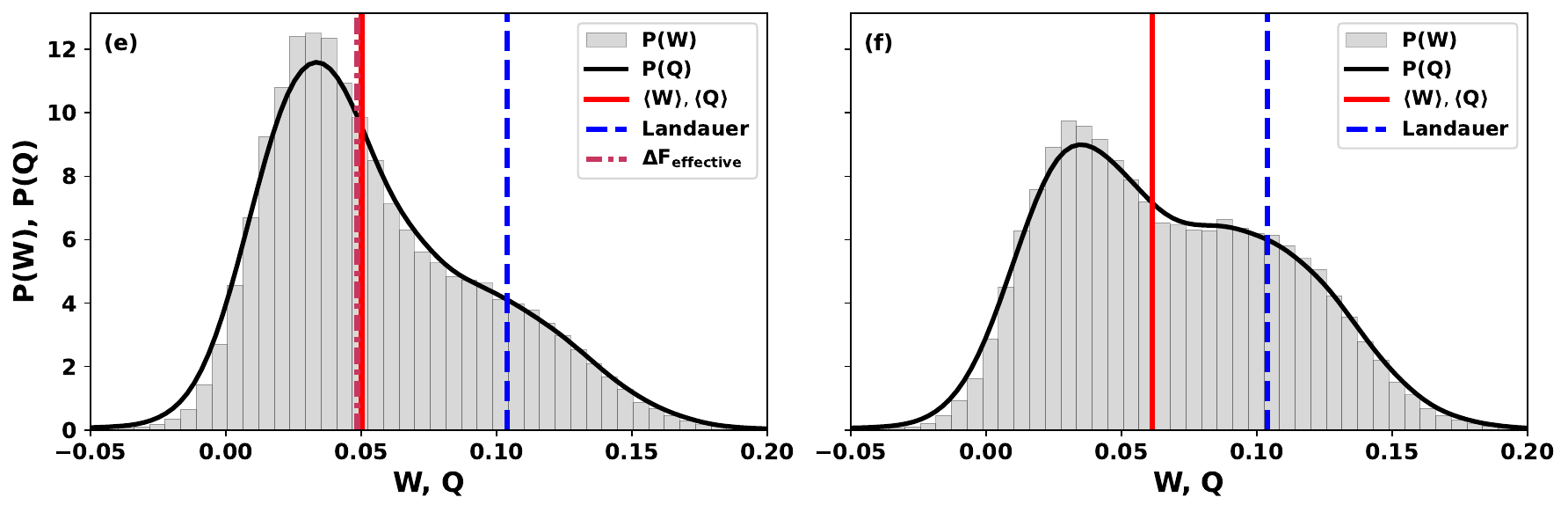}

\caption{
    Work and heat distributions for different asymmetry parameters: (a) and (b) correspond to $c=1$ and $A=1.30$; (c) and (d) correspond to $c=0.74$ and $A=0.90$; (e) and (f) correspond to $c=0.56$ and $A=0.75$. The barrier height of the potential is $h=1$ and the noise strength $D=0.15$. The parameters of the barrier lowering force are $\omega = 0.01$, $Z = 0.5$, The ramp-up and ramp-down durations of the tilting force are $\tau_1 = 298$, and $\tau_2 = 16$, respectively. The results in panels (a), (c), and (e) are for initial equilibrium distributions of the memory states; in panels (b), (d), and (f), we represent the results when particles are initially distributed equally between both wells (50-50 distribution) without thermalization at the start of the erasure protocol. In the sub-figures, the Landauer bound $D\ln 2$ is shown with the dashed line. $\langle W \rangle$ and $\langle Q \rangle$ are indicated with solid straight line and $\Delta F_{\textrm{effective}}$ is presented with dotted dash line.
}
  \label{fig:merged_heat_distributions}
\end{figure*}

As per the design of the erasure process, we suggested that the protocol be regarded as a cycle. The entire probability distributions of work and heat for a full cycle, for complete memory erasure, are examined and shown in Fig.~\ref{fig:merged_heat_distributions} for both symmetric and asymmetric cases. We illustrate the work distribution, $P(W)$ versus $W$, and the heat distribution, $P(Q)$ as a function of $Q$ for the erasure process performed in a symmetric ($c=1$ in Fig.~\ref{fig:merged_heat_distributions}\textcolor{blue!}{(a)-(b)}) and two asymmetric potentials ($c=0.74$ in Fig.~\ref{fig:merged_heat_distributions}\textcolor{blue!}{(c)-(d)}  and $c=0.56$ in Fig.~\ref{fig:merged_heat_distributions}\textcolor{blue!}{(e)-(f)}). For each case, we consider two different starting conditions for the erasure mechanism. First, we examine the erasure protocol with the steady-state distribution of the initial memory states. Then, for the same system, a separate erasure cycle is designed where we begin with an initial 50:50 distribution in both wells and do not provide any thermalization period to the system. In the case of the asymmetric potentials, these two distinct scenarios correspond to the erasure processes starting with the initial equilibrium and nonequilibrium memory states, and the findings lead to the interpretations of the effects of the nature of the initial memory states on the subjects under study. However, for the symmetric potential, these two initial distributions of the memory states are identical, as discussed. This fact is also reflected in the work and heat distribution plots when erasure is done in an intrinsic symmetric potential. The distributions appear indistinguishable for the erasure mechanisms with two different initial conditions mentioned above when the underlying potential has the symmetric form (Fig.~\ref{fig:merged_heat_distributions}\textcolor{blue!}{(a)-(b)}). Now, we describe and analyze the characteristics of the work and heat distributions for both symmetric and different asymmetric cases and with two specific initial conditions for each of them, in more detail. In all cases, we observe bimodal distributions for work and heat involved with the erasure process. The left peak in the distribution curves appears because of the proportions of the particles initially present in the right well or in the memory state ``$1$" before the protocol starts. This peak centers around the low values of $W$ and $Q$ as they essentially correspond to the movements of the particles within the same well. In contrast, the high value peak or the right peak in the work or heat distribution arises due to particles moving from the left well to the right potential well, i.e., for the transitions from the memory state ``$0$'' to ``$1$'' during the erasure process. Although there are similarities in the structures of the distributions for the symmetric and the asymmetric cases, noticeable differences become evident if we concentrate on the quantitative values of $\langle W \rangle$ and $\langle Q \rangle$ obtained from the distributions in these two distinct scenarios. We can observe that the $\langle W \rangle$ and $\langle Q \rangle$ fall below the Landauer bound when asymmetry in the potential is implemented. This becomes apparent when we focus on the Landauer bound, and $\langle W \rangle$ or $\langle Q \rangle$ lines in the work or heat distribution plots and compare their positioning for the symmetric and asymmetric cases. The observations suggest that for the symmetric condition, the values of $\langle W \rangle$ or $\langle Q \rangle$ are above the Landauer bound. However, one can reach below the Landauer limit for the finite-time erasure processes for the asymmetric cases. The decrease of $\langle W \rangle$ or $\langle Q \rangle$ from this limiting value for the asymmetric systems increases with the rise of the degree of asymmetry, i.e., with the decreasing value of $c$. This becomes possible because of the structural design adopted in the asymmetric potential, which ultimately facilitates the erasure process to the desired state. Another important point to note here is that for the system with the same extent of asymmetry, a greater contribution of the ensemble is observed around the low-value peak of the work or heat when we start with the equilibrium initial distribution of the memory states. This suggests a larger occupancy of the target memory state as compared to the initial equal distribution of memory states at the beginning of the erasure protocol. Consequently, the values of $\langle W \rangle$ and $\langle Q \rangle$ are lower for the former case, i.e., more below the Landauer bound, and can be considered more energetically favorable.

\begin{figure}[ht]
  \centering
  \includegraphics[width= 0.98\linewidth,height=6cm]{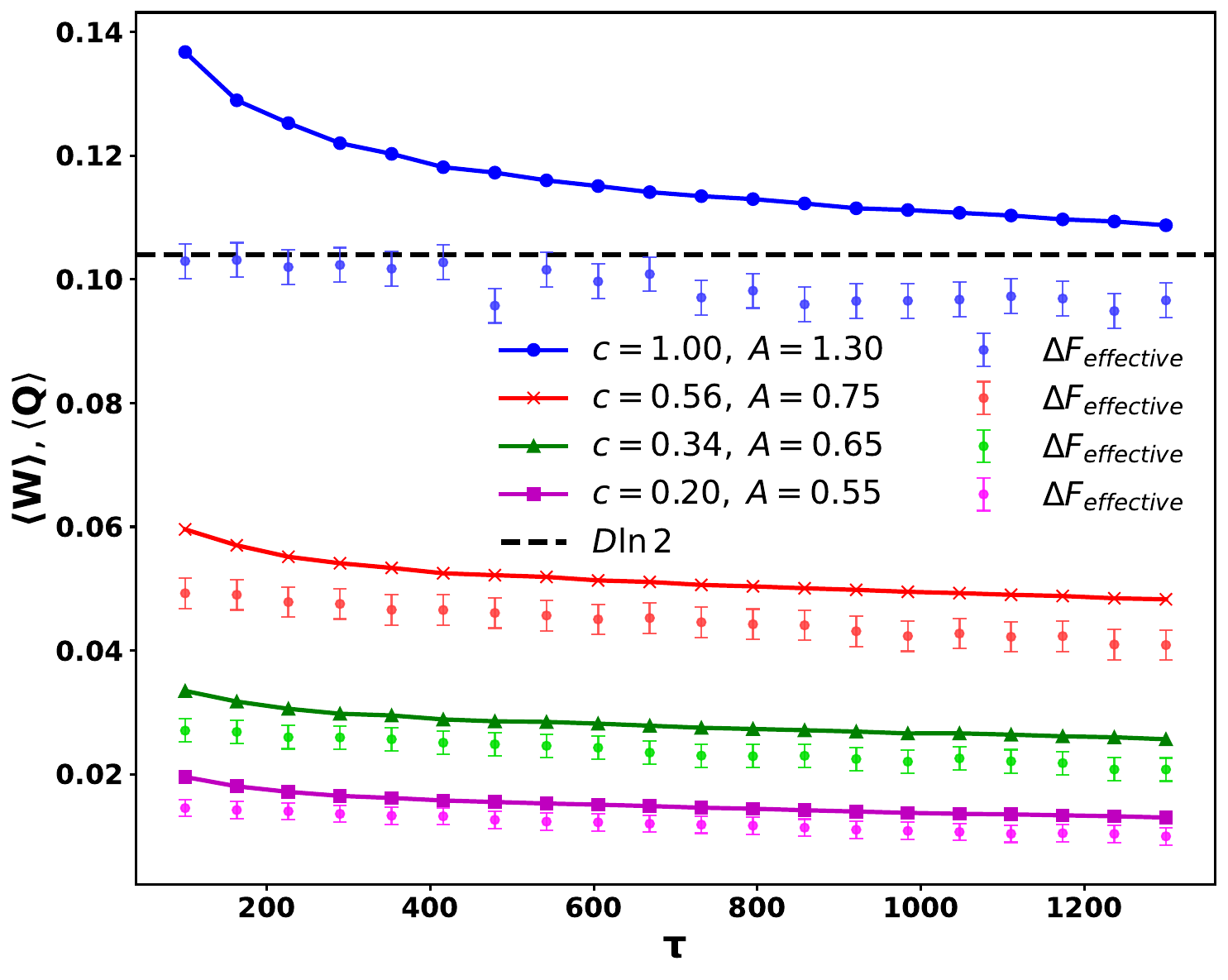}
  \caption{Mean work \(\langle W \rangle\) and heat \(\langle Q \rangle\) versus protocol duration \(\tau\) for a fixed potential height \(h = 1\), noise strength \(D = 0.15\), barrier lowering parameters $\omega = \pi / \tau$, and $z=0.5$. The system is first thermalized for a duration of \(t_0 = 314\), followed by a protocol phase with a ramp-up and ramp-down time ratio of \(\tau_1 : \tau_2 = 19 : 1\) for the tilting force. The results have been presented for four asymmetry parameter values $c=1, 0.56, 0.34, \textrm{and}\;\ 0.2$. The corresponding $A_{\textrm{opt}}$ has been chosen as the amplitude of the tilting force for each case. The estimated values of $\Delta F_{\textrm{effective}}$ with error bars, calculated in the given range of $\tau$, are plotted for all values of $c$, to compare with the corresponding  $\langle W \rangle$ and $\langle Q \rangle$.}
  \label{fig:workonly_0}
\end{figure}

\begin{figure}[ht]
    \centering
    \includegraphics[width= 0.94\linewidth,height=5.5cm]{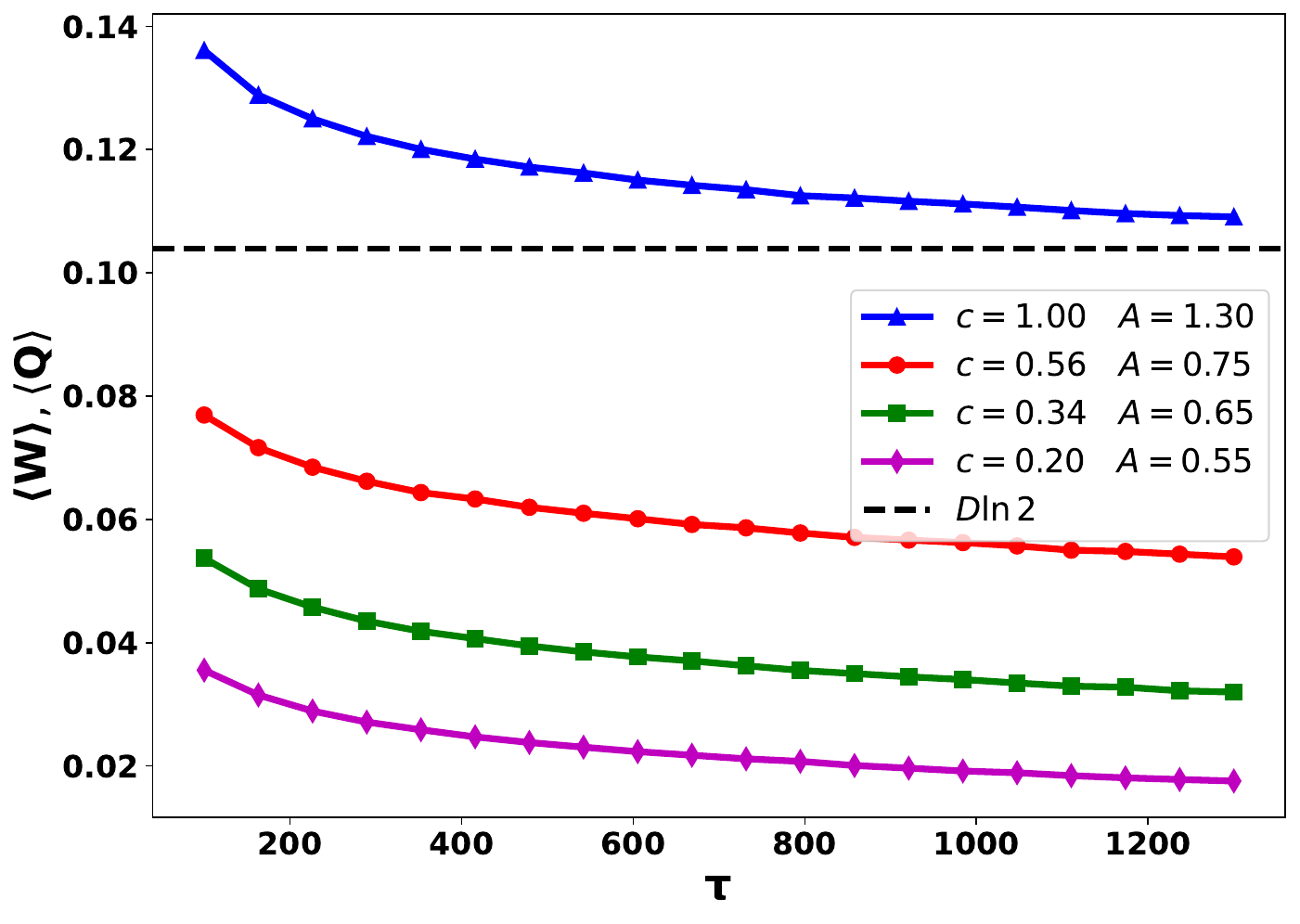}
     \caption{
    Mean work \(\langle W \rangle\) and heat \(\langle Q \rangle\) versus protocol duration \(\tau\) for a fixed potential height \(h = 1\), noise strength \(D = 0.15\), barrier lowering parameters $\omega = \pi / \tau$, and $z=0.5$. The protocol phase has a ramp-up and ramp-down time ratio of  \(\tau_1 : \tau_2 = 19 : 1\) for the tilting force. The results have been presented for four asymmetry parameter values $c=1, 0.56, 0.34, \textrm{and}\;\ 0.2$. The corresponding $A_{\textrm{opt}}$ has been chosen as the amplitude of the tilting force for each case. In all cases, erasure mechanisms start with an initial equal distribution (50:50) of the memory states, and no thermalization is allowed before the start of the protocol. This initial distribution corresponds to the nonequilibrium initial state for all asymmetric systems.
  }
  \label{fig:nonworkf5050}

\end{figure}

\begin{figure}[ht]
  \centering
  \includegraphics[width= 0.97\linewidth,height=5.5cm]{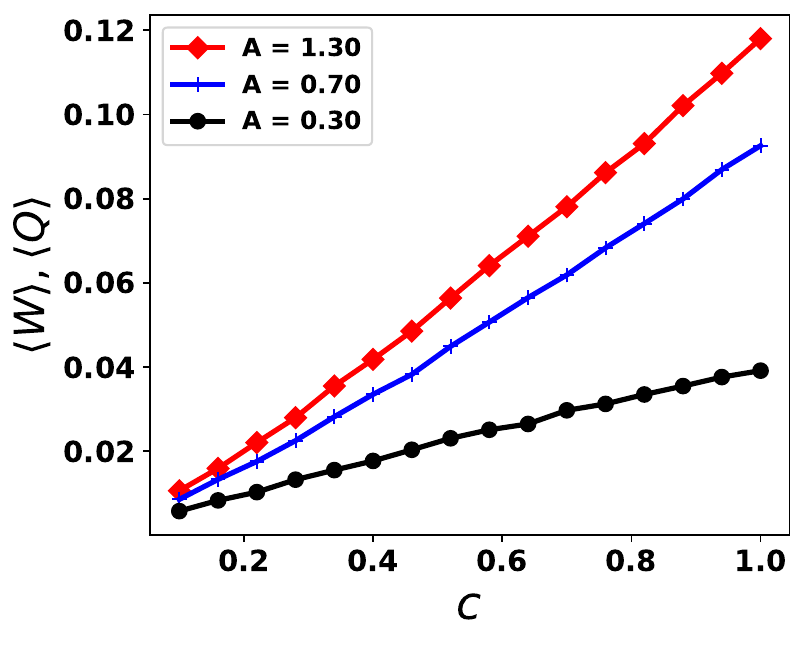}
  \caption{Mean work \(\langle W \rangle\) and heat \(\langle Q \rangle\) versus the asymmetry parameter $c$ for a fixed potential height \(h = 1\), noise strength \(D = 0.15\), barrier lowering parameters $\omega = \pi / \tau$, and $z=0.5$. The system is first thermalized for a duration of \(t_0 = 314\), followed by a protocol phase with a ramp-up and ramp-down time ratio of  \(\tau_1 : \tau_2 = 19 : 1\) for the tilting force. The results have been presented for three values of the tilting force amplitudes.}
  \label{fig:workvsc_0D_0.15}
\end{figure}

\begin{figure}[ht]
  \centering
  \includegraphics[width= 0.97\linewidth,height=5.5cm]{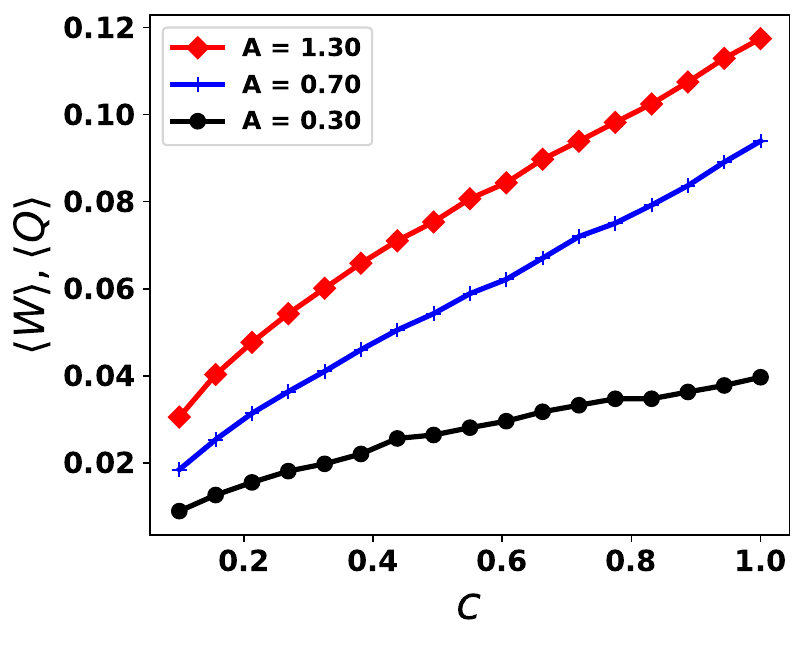}
 \caption{ Mean work \(\langle W \rangle\) and heat \(\langle Q \rangle\) versus the asymmetry parameter $c$ for a fixed potential height \(h = 1\), noise strength \(D = 0.15\), barrier lowering parameters $\omega = \pi / \tau$, and $z=0.5$. The protocol phase has a ramp-up and ramp-down time ratio of  \(\tau_1 : \tau_2 = 19 : 1\) for the tilting force. The results have been presented for three values of the tilting force amplitudes. In all cases, erasure mechanisms start with an initial equal distribution (50:50) of the memory states, and no thermalization is allowed before the start of the protocol. This initial distribution corresponds to the nonequilibrium initial state for all asymmetric systems.}
  \label{fig:workvsc_5050D_0.15}
\end{figure}

Next, we analyzed the mean work \(\langle W \rangle\) and heat \(\langle Q \rangle\) as a function of protocol duration \(\tau\) under the influence of both external driving force and barrier modulation, with \(\tau_1 : \tau_2 = 19:1\). While many experimental studies ~\cite{berut_2013} have neglected the barrier modulation contribution, we explicitly account for both functions in our analysis. It is understood that the average value of the work and heat decreases with the increasing value of $\tau$ as we approach the quasi-reversibility ~\cite{berut_lutz2012nature,moupriya_2014}. This is observed for erasure mechanisms in different cases with the symmetric and asymmetric potentials. Also, it is known that in the very long-time limit, these average quantities advance towards the Landauer bound for the erasure process in the symmetric case \cite{berut_lutz2012nature,Moupriya_PhysRevE_2014,moupriya_2014}. This is what has again been retrieved in our present analysis of erasure for the $c=1$ case. Interestingly, for the asymmetric cases, we observe a violation of the Landauer bound. The results have been shown in Fig.~\ref{fig:workonly_0}. We compare the findings for the equilibrium and nonequilibrium initial memory states. The observations for the nonequilibrium initial memory condition have been presented in Fig.~\ref{fig:nonworkf5050}. We detect that the basic interpretations remain the same for both circumstances. Sagawa and Ueda~\cite{Sagawa_2009} previously demonstrated that Landauer's principle can be violated when the initial state is in nonequilibrium and the system exhibits asymmetry. This theoretical prediction was subsequently verified experimentally by Ciampini et al.~\cite{ciampini2021experimental}. However, in our present study, we also allow the system to reach equilibrium before initiating the protocol, suggesting that asymmetry alone can lead to violations of the Landauer bound even under equilibrium initial conditions. Moreover, we explore in detail how the extent of asymmetry can affect the average values of work and heat associated with the finite-time erasure processes. To develop a more systematic understanding of this fact, we scrutinize the variation of \(\langle W \rangle\) and \(\langle Q \rangle\) with the asymmetry parameter \(c\) for a fixed value of the amplitude of the erasure force $A$. We find that for higher asymmetry, we need to do much less work compared to the symmetric case. The results for three fixed values of $A$ are shown in Fig.~\ref{fig:workvsc_0D_0.15}. The same has been examined for the erasure protocols with nonequilibrium initial conditions of the memory states. The results have been illustrated in Fig.~\ref{fig:workvsc_5050D_0.15}. From the comparisons of the data in Fig.~\ref{fig:workvsc_0D_0.15} and Fig.~\ref{fig:workvsc_5050D_0.15}, we detect a monotonic rise of $\langle W \rangle\ $ and $\langle Q \rangle\ $ as a function of $c$ in both cases. However, the rate of the rise is steeper in the case of the equilibrium initial memory condition.


\subsection{Analysis of the results with the detailed Jarzynski equality}
\begin{figure*}[ht]
        \includegraphics[width=0.49\linewidth,height=5.5cm]{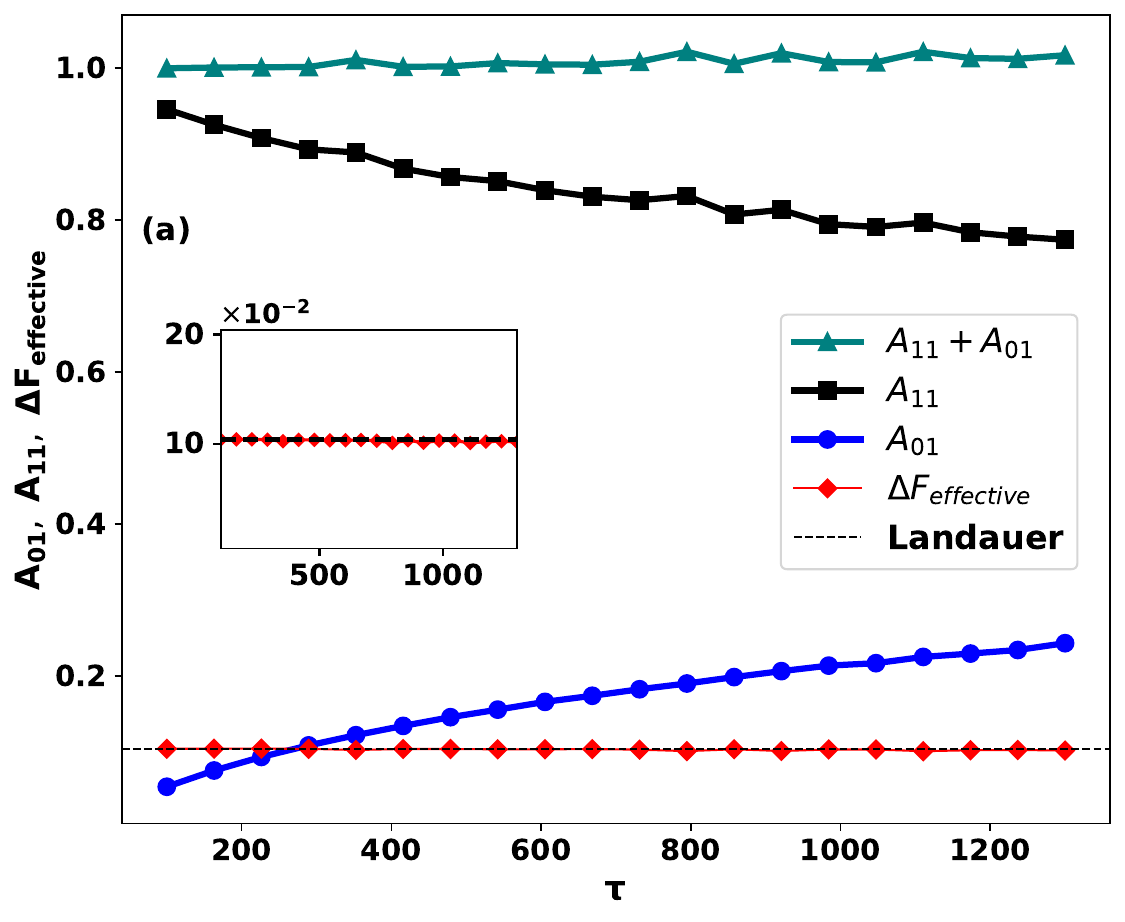}
        \includegraphics[width=0.49\linewidth,height=5.5cm]{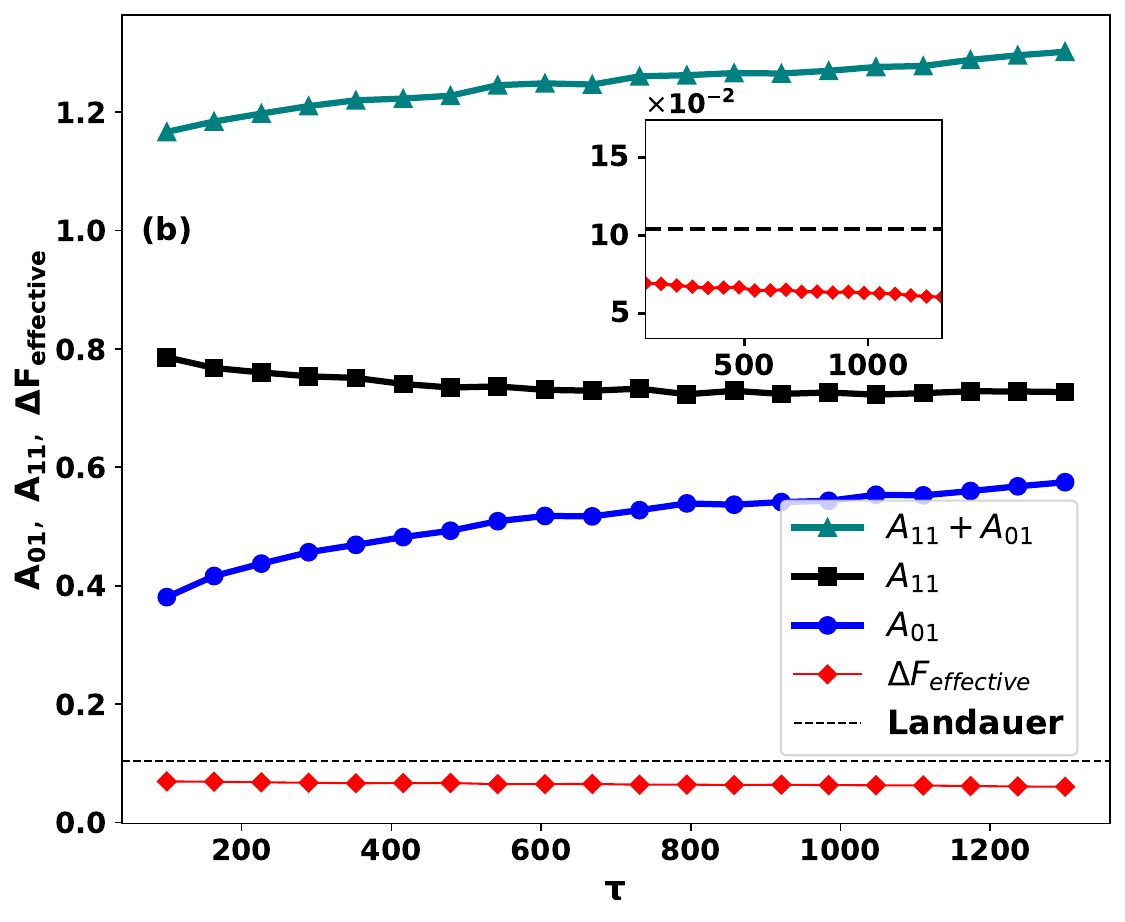}


        \includegraphics[width=0.49\linewidth,height=5.5cm]{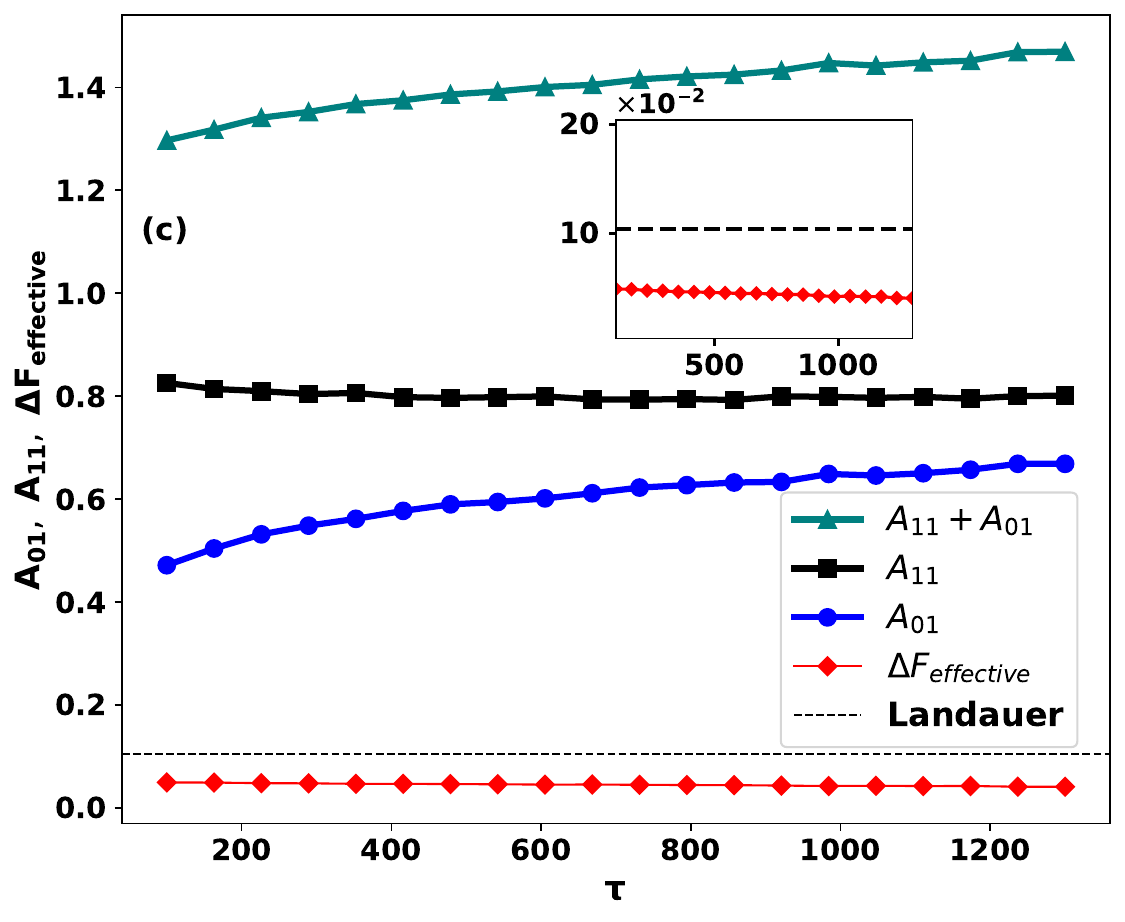}
        \includegraphics[width=0.49\linewidth,height=5.5cm]{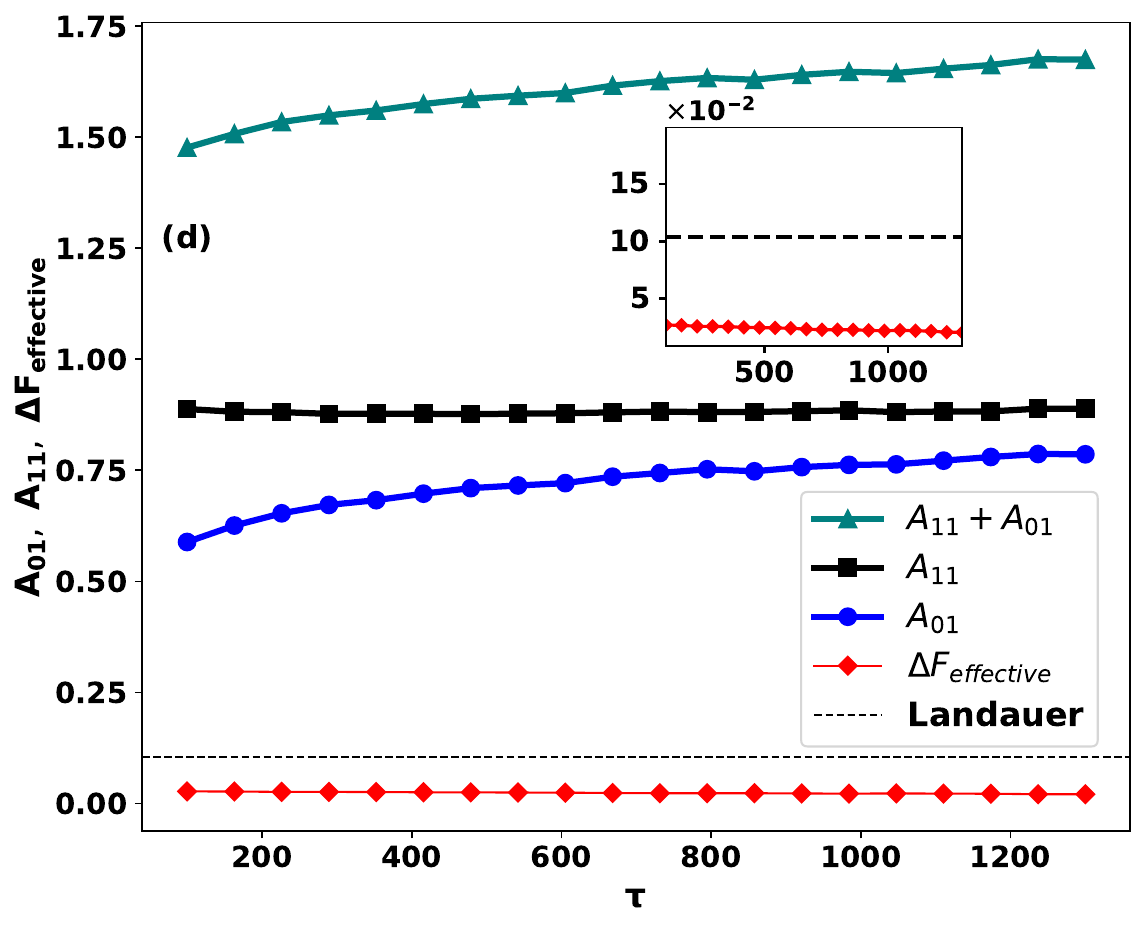}
   
        \includegraphics[width=0.49\linewidth,height=5.5cm]{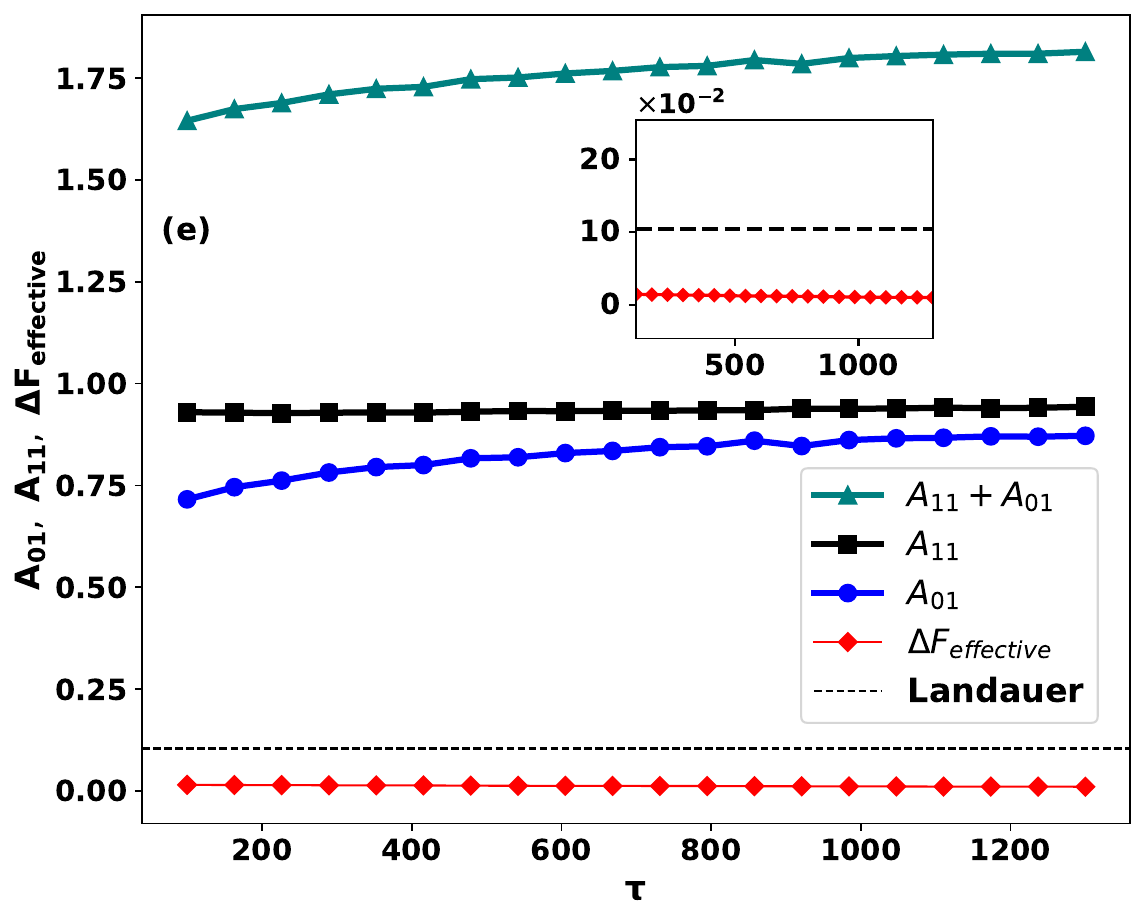}

    \caption{
        Plots of \(A_{01}\), \(A_{11}\), \(A_{01}+A_{11}\) and \(\Delta F_{\textrm{effective}}\) versus \(\tau\) for five different values of \(c\) (\(c = 1\), \(c = 0.74\), \(c = 0.56\), \(c = 0.34\), and \(c = 0.20\)), with fixed barrier height \(h = 1\), noise strength \(D = 0.15\), and barrier lowering parameters $\omega= \pi/ \tau$, and \(Z = 0.5\). The erasure cycles follow a ramp-up and ramp-down time ratio of \(\tau_1 : \tau_2 = 19 : 1\) for the tilting force. The corresponding amplitude of the tilting force are $A=5.0, 0.9, 0.75, 0.65,$ and $0.55$, respectively. In all cases, the erasure protocols start with the initial equilibrium distribution of the memory states.
    }
    \label{fig:free_c}
\end{figure*}

\begin{figure}[ht]
 \centering
 \includegraphics[width= 0.97\linewidth,height=5.5cm]{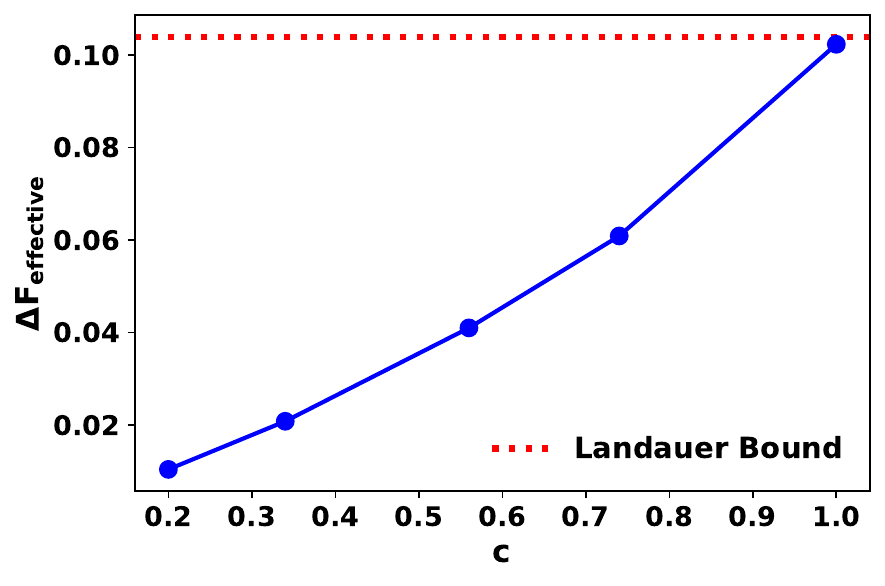}
 \caption{Plot of \(\Delta F_{\textrm{effective}}\) versus \(c\) at a fixed \(\tau = 1300\), with a constant barrier height \(h = 1\), noise strength \(D = 0.15\), and barrier lowering parameters $\omega=\pi/ \tau$, and \(Z = 0.5\). The erasure cycles follow a ramp-up and ramp-down time ratio of \(\tau_1 : \tau_2 = 19 : 1\) for the tilting force. In all cases, the erasure protocols start with the initial equilibrium distribution of the memory states.}
  \label{fig:freevsc}
\end{figure}

We realize that in the classical set-up of the erasure mechanism in a symmetric bistable potential, the Landauer bound for the erasure of a classical bit of information is approached when the cycles are extremely slow, i.e., $\tau \to \infty$ ~\cite{berut_2013,moupriya_2014,Moupriya_PhysRevE_2014}. However, it is understood that if the free energy change ($\Delta F$) for the process is estimated, the Landauer bound can be verified in terms of $\Delta F$ irrespective of the cycle duration $\tau$ ~\cite{berut_2013,Buffoni_pre}. This is because, according to the thermodynamic principles, the free energy change corresponds to the lower limit of the work done for any process, in general. Now, the main question is how to get the difference in free energy between the initial and final states of the system for the erasure mechanism. To answer this question, we need to implement the Jarzynski equality, which allows us to calculate the free energy change between two states of a system in contact with a heat bath at temperature $T$. This measurement is done by employing the distribution of the work that is involved in the transformation between the states through nonequilibrium paths. The Jarzynski equality connects the change in free energy during any nonequilibrium process, in general, \(\Delta F\), with the work done, \(W\)~\cite{Jarzynski_prl_1997,Jarzynski_pre_1997,ChrisJarzynski_2004,sagwa_prl_jarjyanski_2010,buffoni2022spontaneous,pre_Williamson_2024}, when both the initial and the final states are at equilibrium. The Jarzynski relation is expressed as follows,
\begin{equation}\label{3.5}
    \langle e^{-\beta W}\rangle = e^{-\beta \Delta F}
\end{equation}
where \(\beta =1/k_{B} T  \). Now, we proceed to understand the free energy change for the erasure process through the application of the Jarzynski equality. For a complete erasure cycle, $f(t=0) = 0$ and $f(t_0 + t_f) = f(t_0) = 0$. As this is a cyclic process, there will be no free energy change between the initial and final equilibrium states of the system, $F$ being a state function. i.e., $\Delta F = 0$ or $\langle e^{-\beta W}\rangle = 1$. This points towards the fact that the classical Jarzynski equality cannot be used to retrieve the Landauer bound in terms of the free energy change. This difficulty arises as the final state of the successful erasure cycle differs from the final equilibrium state. To estimate the free energy change from the work effects under these conditions, we need to consider the detailed Jarzynski equality~\cite{berut_2013,Kawai_2007, Vaikuntanathan_2009}, which links the free energy change and nonequilibrium work when the final state is significantly different from the equilibrium state. We define an effective free energy change for the memory erasure process in which the memory ends in a unique desired state, as \(\Delta F_{\text{effective}}\). This effective free energy is related to the nonequilibrium work through the relation~\cite{moupriya_2014,Moupriya_PhysRevE_2014}
\begin{equation}\label{3.6}
    \langle e^{-\beta W}\rangle _{\to 1}  = e^{-\beta \Delta F_{\textrm{effective}}}
\end{equation}
where \(\langle \cdot \rangle_{\to 1}\) denotes the average taken over trajectories that end in the target state at the end of the erasure cycle. For our present setup, this desired final state is the right well or the memory state $1$. We divide this into two sub-processes, as in ~\cite{berut_2013,moupriya_2014}: one process being the transitions of particles from the left to the right well, and the other considers the movements of the particles in the right well, which were already there at the start of the erasure protocol. In other words, the first sub-process can be described as the transition from the memory state $0$ to $1$ during the erasure cycle, and the second one defines the scenario where the memory state remains unaltered as $1$ at the beginning and at the end of the erasure protocol. The values of the average exponential work functions for these two sub-processes are defined as:
\begin{equation}\label{3.7}
    A_{01} = \langle e^{-\beta W} \rangle_{0 \to 1}, \quad A_{11} = \langle e^{-\beta W} \rangle_{1 \to 1}
\end{equation}
where \(\langle \cdot \rangle_{i \to j}\) is the average value of the function for trajectories of particles that initially start in the \(i\)th memory state and end in the \(j\)th memory state after the erasure protocol. With this definition, Eq.~\ref{3.6} can be rewritten as~\cite{moupriya_2014},
\begin{eqnarray}\label{3.8}
e^{-\beta \Delta F_{\textrm{effective}}} = \langle e^{-\beta W}\rangle _{\to 1} = \dfrac{A_{01}+A_{11}}{2}. 
\end{eqnarray}
The factor \(1/2\) appears because of equal distribution in both wells after the thermalization. We have plotted \(A_{01}\), \(A_{11}\), and \(\Delta F_{\textrm{effective}}\) for various asymmetry parameters \((c)\) as functions of $\tau$. We found that for the symmetric case \((c=1)\), the sum of \(A_{01} + A_{11}\) is close to \(1\), which has already been reported~\cite{berut_2013,moupriya_2014}, and this value corresponds to \( \langle e^{-\beta W} \rangle_{\to 1} = 1/2\), i.e., the free energy change is \(\Delta F_{\text{effective}} = D \ln 2\). This establishes the fact that the Landauer bound can be recovered in terms of free energy change. The results have been shown in Fig.~\ref{fig:free_c}{\textcolor{blue}{(a)}}. However, we observed that in the asymmetric cases, the sum is always greater than \(1\) for all degrees of asymmetry, because the system always tries to maximize the entropy and prefers the wider potential well. Therefore, the free energy calculation is
\begin{align}\label{3.8b}
\Delta F_{\text{effective}} &= -D \cdot \ln \left( p_{01} A_{01} + p_{11} A_{11} \right) 
\end{align}
where $p_{01} = \frac{N_{01}}{N_{01} + N_{11}}$ and $p_{11} = \frac{N_{11}}{N_{01} + N_{11}}$ are the relative fractions of each type of transition, conditioned on the final memory state being ``1''. The free energy change falls below the Landauer bound, as shown in Figs.~\ref{fig:free_c}{\textcolor{blue}{(b)-(e)}} for four different asymmetric cases studied. It is noticed that the value of the estimated $\Delta F_{\textrm{effective}}$ is specific for each value of $c$. It increases with increasing $c$ and ultimately converges to the Landauer bound for the symmetric potential with $c=1$. These findings have been demonstrated in Fig.~\ref{fig:freevsc} for a specific value of the erasure cycle duration $\tau$. Now, very importantly, we identify that the $\langle W \rangle$ or $\langle Q \rangle$ approaches the calculated value of the corresponding $\Delta F_{\textrm{effective}}$ for larger durations of the erasure cycles, irrespective of the value of the asymmetry parameter $c$ (Fig.\ref{fig:workonly_0}). It is also observed in this plot (Fig.\ref{fig:workonly_0}) that the values of the $\Delta F_{\textrm{effective}}$ are more or less constant, for all values of $\tau$, with insignificant fluctuations, as expected, for a given value of $c$. This allows us to establish a bound on the evolved heat or associated work for the bit erasure process for asymmetric systems as well. We also compare $\langle W \rangle$ or $\langle Q \rangle$ with $\Delta F_{\textrm{effective}}$ in the probability distribution plots of $W$ and $Q$ for some representative values of $c$ (Figs.\ref{fig:merged_heat_distributions}(a), (c), and (e)). The decreasing magnitudes of this limiting value, interpreted in terms of $\Delta F_{\textrm{effective}}$, for higher degrees of asymmetry in the potential, establish the fact that the erasure protocols designed with asymmetric potentials can achieve lower thermodynamic costs. Our systematic analyses develop a quantitative understanding of this advantage that can be regulated through the tuning of the degree of asymmetry of the underlying potential. We emphasize that the outcomes of this particular investigation regarding $\Delta F_{\textrm{effective}}$ appear to be very significant. We make this remark as this exploration allows us to suggest a limit for the average values of the work or heat involvement for the erasure processes in small systems when the Landauer bound is supposed not to hold due to the intrinsic asymmetry of the governing potential. 

There is one crucial aspect to mention at this point. In the case of the interpretation of the lower limit of the work or heat involvement in terms of the effective free energy change for the erasure mechanism through the application of the detailed Jarzynski equality, we consider the equilibrium distribution of the initial memory states. This has been done for all systems under study, symmetric and asymmetric systems, with varied levels of asymmetries. This is because the implementation of the detailed Jarzynski equality will be valid even if the final state is a nonequilibrium state; however, the initial state has to be at equilibrium. The effective free energy change cannot be defined if both the initial and final states are not at equilibrium. Therefore, we depict the analyses of the effect of the asymmetry of the intrinsic potential on the effective free energy change for the erasure process, where we consider equilibrium initial conditions for all cases examined. As the calculation of the effective free energy bound is not possible for all cases (with initial nonequilibrium distributions of the memory states), we preserve the Landauer bound line as a reference for comparison with the $\langle W \rangle$ or $\langle Q \rangle$ in Figs.\ref{fig:merged_heat_distributions}, \ref{fig:workonly_0}, and \ref{fig:nonworkf5050}. Although the calculations of $\Delta F_{\textrm{effective}}$ are not possible for the cases with nonequilibrium distributions of the initial memory states, we can comment on considering this quantity as a general lower bound for $\langle W \rangle$ or $\langle Q \rangle$ for the erasure processes regardless of the initial conditions. This is because when we scrutinize the values of $\langle W \rangle$ or $\langle Q \rangle$ for two distinct initial distributions, keeping all other factors the same, we find that the values are slightly higher for the nonequilibrium starting condition cases. The corresponding values become equal only for the symmetric system, as in that case, the initial equilibrium distribution and a 50:50 distribution of the memory states are identical. Therefore, we validate that the values of $\langle W \rangle$ or $\langle Q \rangle$ do not reach below $\Delta F_{\textrm{effective}}$ in any cases for the erasure mechanisms under study. Consequently, we can say that the quantity sets a general lower limit for the work or heat of erasure of a classical bit of information when the initial memory states can have asymmetric as well as nonequilibrium characteristics.

\subsection{Validity in the parameter regime}
Following the thorough analyses of the erasure process, exploiting the 
asymmetry of the governing potential, through the success rate, work (heat), 
and free energy calculations, we now explicitly discuss the justification 
of the choice of parameters in these studies. This will give a definite idea 
about the range of parameter values, linked to the system and dynamics, for which the above interpretations will be valid.

First, we discuss the rationale regarding the selected parameter values 
in the stage of the preparation of the memory states before the start of the 
erasure protocol. Throughout the study, the time-independent barrier height 
of the potential has been kept fixed at $h=1$, 
irrespective of the value of the asymmetry parameter $c$. In most cases, 
the value of $D$ is considered to be $0.15$. Therefore, compared to the barrier height, the noise 
strength satisfies the condition of a weak noise regime for the systems under study. 
This property of the environmental noise is required to certify the stability 
of the designed memory states. As the noise strength is sufficiently low compared to the 
barrier height of the potential, the transitions between the two wells, 
i.e., between the two memory states, are rare. This aspect takes care of the fact that the 
two types of memory states are well-defined, separated by a significant magnitude of the 
energy barrier. In the case of the studies with the initial thermalized distributions, 
the systems are first allowed to get equilibrated with the heat bath for a certain range of time.
This time span is adequate for the systems to reach their steady state distributions 
as no significant changes in the distributions are observed if the systems evolve 
for a longer time. This fact has been verified for the system with the symmetric potential
as well as for all asymmetric systems with a range of $c$ values considered in our studies. 
This examination suggests the robustness of the initial thermalized distribution and, 
consequently, the preparation of the memory states at the start of the protocol 
for the concerned cases. For the analyses with initial nonequilibrium distributions, 
we start with the $50:50$ distribution of the memory states for 
the systems with all values of $c$. In this case, we do not let the system get thermalized with 
the heat bath. Therefore, the thermal noise does not affect the preparation 
of the memory states for these circumstances.

Next, we scrutinize the chosen parameter values during the operation of the 
erasure cycles, to remark on the reliability of the selection of their magnitudes 
in constructing apt erasure mechanisms. The parameter $Z$ present in the barrier-lowering force 
(Eq.~\ref{eq:barrier_lowering}) determines the minimum height of the potential barrier during the erasure mechanism. 
As its value has been considered to be equal to $0.5$ for all cases studied, 
the lowest value that the barrier height can attain is $0.5$. Therefore, we suggest  
that a reasonable value of the barrier height to noise ratio (greater than $3$, for our 
specific choice of parameters) is maintained 
even for the situation with the least barrier height. This implies that the effect of 
the noise will not be dominant in the overall dynamics, 
so that the roles of the underlying potential and the erasure forces 
remain obscured during the process. At the same time, we mention that 
the barrier-lowering protocol is implemented in the erasure mechanism 
to allow for a greater amount of thermally activated transfer between the two states 
as compared to the unperturbed system~\cite{berut_lutz2012nature, Dillenschneider_prl_2009}. 
In the latter case, the distinction between the two memory states is more rigid 
and requires a stricter restriction on the observed overall stability of the two states. 
With reference to these facts, we point out that the required conditions are maintained 
throughout our studies, both for the initial conditions and during the erasure protocol. 
This ultimately verifies that the resulting erasure phenomenon captures the  
integrated effect of the underlying potential, environmental noise 
and the erasure forces appropriately. 
We also discuss the choice of the value of $\omega$, which is related to the tilting force 
time period $\tau$ for finite-time erasure as $\pi/ \omega$. 
This consideration guarantees that the barrier-lowering force arrives 
at its initial value $0$ at the end of the cycle, following the appropriate protocol. 
In the majority of the studies, the value of $\omega$ has been taken to be equal to one, which 
corresponds to $\tau=314$. This value of $\tau$ is comparable to the thermal relaxation 
time-scale of the systems, which has been discussed in the context of the initial thermalization 
of the two memory states. We also performed studies for varying values of $\tau$ and 
the magnitude of $\omega$ has been defined accordingly.

Finally, we indicate that the observed variation of the success rate versus $D$ plot 
(Fig.\ref{fig:e_d_c_0_position}(a)) is very informative to comment on the validation of the 
considered value of $D$ in comparison to the barrier height $h$. 
We detect that the success rate remains above a high threshold value (approximately above $95\%$) 
for all values of $c$, i.e., for all asymmetric and symmetric systems. 
This indicates that the erasure process works efficiently within this range of the 
noise strength for the class of systems under study. As we perform most of our analyses, 
considering $D$ value equal to $0.15$, we suggest that the results obtained from the present 
study correspond to effective erasure mechanisms. Also, we imply that the overall interpretation 
of the present work regarding the improvements of the erasure process in terms of both kinetic 
and thermodynamic factors will be valid within this weak noise range.

\section{Conclusion and Outlook}
In our current work, we focus on some crucial aspects 
linked to the fundamental process of information erasure 
where the information is stored as the 
classical bits of binary memory, $0$ and $1$.  
The novel aim of the study 
is to explore the possibility of improving the performance 
of the erasure process, both in terms of the success rate 
and the amount of essential heat evolution and work-done  
associated with it. These are the critical factors that contribute to 
the overall efficiency with which tasks are carried out 
in digital devices as the erasure steps are indispensable 
for their necessary operations. 
The goal is to increase the success rate and 
to decrease the extent of heat released to the environment,
for the erasure mechanism, which would effectively 
improve the performance of the devices.  
The limit for the evolved heat for the erasure of a 
classical bit of memory is set by the Landauer bound of 
$k_{B}T \ln2$, which is approached for an erasure protocol of 
infinite duration. A greater amount of heat is released for 
any finite-time erasure process. In the case of the small 
systems, the same description applies to the average of the 
quantity, heat~\cite{Dillenschneider_prl_2009}. This is understood through 
the design of the erasure mechanism when the two memory states 
are modeled in a symmetric bistable potential, and 
they are separated by a symmetric barrier. 
We address the objectives of the present work, 
by carrying out a systematic analysis 
of modulating the degree of asymmetry of the underlying 
potential barrier separating the two binary memory states $0$ and $1$. 
This asymmetry is implemented through different widths of 
the two potential wells of the bistable system  
and the asymmetric shape of the potential barrier 
partitioning them, although the two minima 
of the potential wells representing the memory states 
are energetically equivalent. 
For this purpose, we consider the potential as described in Eq.~\ref{2.2}. 
Despite the distinct mathematical form of this bistable potential 
and the particular parametrization, the potential function for asymmetric cases 
can account for the generic asymmetry in the dynamics through the introduction 
of an entropic bias between the two states, 
whereas keeping their energies similar ~\cite{innerbichler_2020}. 
Therefore, this choice of potential guarantees the energy equivalence of the 
two binary memory states. Also, at the same time, it allows us to investigate the 
effect of the asymmetry of the governing potential on the erasure process, in general.
We systematically scrutinize this aspect of the asymmetry 
in the potential form to find out any significant impact 
on the improvements of the effectiveness of the erasure protocol.

It has been found recently~\cite{Chupeau_2020,innerbichler_2020} that the 
rate of the diffusive transport 
can be increased by optimized tuning of the 
potential barrier shape, 
which modulates the potential to different degrees of asymmetry. 
The key observation is that the 
speed of the barrier crossing process enhances 
for the asymmetric barrier shapes as 
compared to the symmetric case when 
all other conditions remain the same. 
Also, a larger extent of asymmetry has 
greater significance on the increase in the rate. 
We proceed to investigate whether this 
fundamental finding on the transition processes 
influences the erasure phenomenon
that can be partly viewed as a driven barrier-crossing mechanism. 
This is because the proportion of the memories that were initially 
at a different state other than the target state, 
are required to be transferred to the desired state 
at the end of the erasure process. In our physical setup,  
this is realized as the movement of the Brownian particles 
to the right well (memory state $1$), 
if they were primarily present at the left well (memory state $0$),  
to account for the successful erasure. 
Our study reveals that the erasure process becomes 
more favorable in terms of the success rate 
when the potential is more asymmetric in nature. 
To explain, there is a minimum amplitude of the 
erasure force, i.e., of the external linear drive 
which allows for a desired threshold of the 
success rate ($\sim 99\%$) for the erasure process. 
This optimum value of the forcing amplitude 
has a lesser magnitude when the degree of asymmetry 
in the potential is higher. This implies that 
the target level of the success rate 
can be achieved by the application of a lower amount 
of the external force making the process 
feasible under moderate input conditions for asymmetric cases. 
This result can be considered as one of 
the significant outcomes of the present analysis 
as it indicates a direct and methodical way 
of increasing the success rate of the process 
by raising the measure of asymmetry 
in the potential. The increased success rate 
resulting from the facilitated transition 
over the barrier in the appropriate cases, 
can be regarded as the improved kinetics of the process 
in asymmetric potential setups. 
To depict the effect of this asymmetry 
on the energetic aspects, we concentrate on 
the relevant thermodynamic quantities, 
such as the average work done, evolved heat 
and the effective free energy change related to the erasure process. 
We compare the crucial aspects of the investigations 
to analyze the effect of the equilibrium and nonequilibrium distributions 
of the initial memory states on them, wherever applicable. 
As the erasure protocols are set up as cyclic processes, 
i.e., the initial and final values of the external forces 
possess the same value, the numerical value of the average work done becomes 
equivalent to the average heat evolution for erasure 
following the thermodynamic principles. 
Our study leads to a very crucial understanding that 
the average work done or the released heat 
can reach below the Landauer bound 
for the finite-time erasure processes 
under pertinent conditions in the case of the 
asymmetric setups as described. This has also been justified 
by the calculated effective free energy change, which defines the 
lower limit of the average work done, through the 
application of the detailed Jarzynski equality. 
The estimation of this effective free energy change, 
with the aid of the detailed Jarzynski equality, leads us to 
define a general bound for the average work done or evolved heat 
for the erasure processes linked to asymmetric systems. This finding is important 
as we identify a quantity, $\Delta F_{\textrm{effective}}$, which provides an idea  
about the limiting value of $\langle W \rangle$ or $\langle Q \rangle$ when the Landauer bound 
does not hold due to the asymmetric nature of the memory states.
Here, we point out that deviation from the Landauer bound in the case 
of the asymmetric memory has been reported 
before~\cite{Konopik_2020_epl,Gavrilov_2016,talukdar2018designing} 
and our work supports the previous findings. 
Moreover, we perform and present a detailed analysis through which 
the departure from the Landauer bound can be understood quantitatively 
with respect to the degree of asymmetry of the system. Our thorough study 
results in establishing the effective free energy change as the lower limit 
for the average work done and heat released for the erasure mechanism. 
This becomes one of the important measures which can be employed to interpret 
the comprehensive variations of the energetic aspects of erasure of information 
for asymmetric systems as compared to the idealized case with symmetric memories. 

Now, we express our views on the applicability and interpretations 
of the above-mentioned results for the erasure process in a general asymmetric framework. 
As our choice of the potential can validate the results 
of earlier studies, both in the theoretical and experimental cases~\cite{Sagawa_2009,Gavrilov_2016},  
we suggest that the considered potential function can be regarded 
to represent a general setup for the described type of asymmetry. 
The advantage of the present mathematical form of the 
potential is that the degree of asymmetry can be directly 
tuned through the asymmetry parameter $c$. 
The work~\cite{Chupeau_2020} suggests that it is 
possible to understand the actions of the specified type of 
asymmetric potential through the design of experiments 
involving intricate tools of microfluidics and optical tweezers. 
Therefore, there remains further scope for verifying our numerical results 
on the kinetic and thermodynamic aspects of erasure of information in asymmetric systems.  
Here, the mentioned asymmetry characterizes the different phase-space volumes of the 
two types of binary memory states involved in the erasure mechanism.

Here, we refer to that the originality of our present study 
lies in the fact that we provide a thorough quantitative 
interpretation of the improved outputs for the erasure processes 
designed in appropriate asymmetric setups. 
This advancement is realized from both the perspective of 
kinetics; in terms of the success rate, and energy costs; 
estimated through $\langle W \rangle$ or $\langle Q \rangle$, 
and $\Delta F_{\textrm{effective}}$. 
Our thorough numerical study quantifies the deviation from 
the Landauer bound for $\langle W \rangle$ or $\langle Q \rangle$ 
depending on the degree of asymmetry 
in the underlying potential. The detailed analyses lead to 
the assignment of a general lower bound as $\Delta F_{\textrm{effective}}$ 
for the average work or released heat associated with the erasure mechanisms. 
This outcome of the study bears a powerful significance. 
This is because it allows us to predict a limiting value for 
$\langle W \rangle$ or $\langle Q \rangle$ for the cases even when 
the classical Landauer bound does not hold.  
This departure from the Landauer limit can happen due to the non-idealized nature 
of the memory states, as in our present cases. 
We emphasize that it is possible to propose appropriate designs of the 
systems that can improve the operations of the erasure mechanisms 
in terms of the thermodynamic costs 
when the initial memory states have an asymmetric equilibrium or 
nonequilibrium origin. However, we retrieve the Landauer bound 
under the conventional setup with initially 
symmetric and equal distribution of memory states. 
This comparative study gives us the scope of proper quantitative 
understanding of improving the suggested erasure protocols from the 
perspective of energetics. 
Also, our analyses unveil 
another very vital fact that the success of the 
erasure process can be enhanced through the 
tuning of the asymmetry factor governing the dynamics. 
Therefore, it can be considered that 
the current work defines a systematic path 
and develops a detailed comprehension 
for the improvement of the efficiency of the 
fundamental process of erasure. 
We believe that the knowledge acquired from the 
present study would guide the design of more efficient erasure mechanisms. 
Here, we also mention that the analyses and the results of the 
current work suggest the possible ways of improvements for the erasure process, specifically. 
We can expect that the construction of improved erasure mechanisms may upgrade the
overall performance of the digital devices. This is because erasure is an essential step for 
irreversible computation, measurement (resetting to the initial memory state step), 
and communication, the primary functions in digital devices~\cite{landauer_nature_1998}. 
However, there are several other steps involved 
in addition to the erasure mechanism in these three basic operations 
(computation, measurement and communication) occurring inside the digital apparatus. 
Therefore, we indicate that our present study cannot conclusively 
suggest the complete upgradation of the performance of the digital machines, as 
there are many other factors that influence the functions of such devices. 
However, it definitely offers a useful scheme, employing asymmetric memory states, 
for the improvement of the functions of one common essential step, 
the erasure mechanism, which is involved in all three fundamental tasks in digital devices. 
We also point out that the focus on this particular step is very important, as this 
accounts for the heat release in the environment during the devices' actions. 
However, to obtain an integrated understanding of the overall level of execution of the devices,  
precise scrutiny of the performance of each individual step in the 
information processing systems, considering asymmetric memory states, will be required. 
These explorations can be considered as extensive points of research in future 
to aim at designing improved technological devices.  

\section*{Acknowledgments}
V.R. acknowledges IIT Mandi for a fellowship. 
M.D. thanks SERB (Project No. SRG/2022/000296), 
Department of Science and Technology, Government of India, 
and IIT Mandi (Seed Grant No. IITM/SG/MUD/91) for financial support.
The High Performance Computing Cluster facility and Param Himalaya Supercomputing 
facility managed by IIT Mandi are also acknowledged. 
The authors thank Dr. Harsh Soni for insightful discussions.\\\\
\textbf{Author contributions:} M.D. conceptualized and designed the research problem.
V.R. executed the research work. V.R. and M.D. analyzed the results. 
V.R. and M.D. prepared and finalized the manuscript draft.\\\\
The authors express no conflicts of interest.

\bibliographystyle{apsrev4-2-titles} 
\bibliography{reference}

\end{document}